\documentclass[%
 reprint,
 amsmath,amssymb,
 aps,
  prb,
]{revtex4-2}

\usepackage{graphicx}
\usepackage{dcolumn}
\usepackage{bm}
\usepackage{booktabs} 

\begin{document}

\title{Anharmonic Effects in Ge$_2$Sb$_2$Te$_5$ and Consequences on Thermodynamic Stability}

\author{Owain T. Beynon}

\author{Adham Hashibon}%
 \email{a.hashibon@ucl.ac.uk}
\affiliation{Insitute for Materials Discovery, University College London, London, WC1E 6AE, United Kingdom}

\begin{abstract}

Chalcogenides  are an important class of phase change material (PCMs) due to their application in digital memory solutions. Owing to their ability to reversibly cycle between crystalline and amorphous states, their use as phase change random access memory (PCRAM) is of interest, and of the many chalcogenide materials Ge$_2$Sb$_2$Te$_5$ (GST) is a promising candidate owing to its stability and low crystallization temperature. GST possesses two stable crystalline polymorphs, cubic and hexagonal. Studies show that phenomena such as heat transport and thermal lattice expansion drive the phase-change nature of these materials. These phenomena are not incorporated in the harmonic approximation, which is a popular model for describing vibrations in solids. Through \textit{ab initio}  density functional theory (DFT), we computationally investigate the anharmonic behaviour of pristine hexagonal GST, \textit{i.e.} without vacancies or defects, while considering the various stacking models that exist and inclusion of van der Waals (vdW) interactions in our modelling. We present the vibrational analysis of different stacking models in GST; Petrov and Kooi-De Hosson (KDH) models and the quantification anharmonic behaviour. Our calculations find that the KDH model is the most stable stacking sequence, being 88 meV more stable than the Petrov model when considering anharmonicity, where this difference is underestimated using a purely harmonic framework (65 meV). These results demonstrate the importance of incorporating anharmonic and dispersion effects when modelling GST, especially in the  choice of stacking models, along with implications for phenomena relating to phase-change behaviour. \\

\end{abstract}


\maketitle

\section{Introduction}

Chalcogenide GeSbTe is a phase-change material useful for applications as novel non-volatile memory in electronics \cite{yamada_high_1987, kolobov_possible_2024, wuttig_phase-change_2007}. Phase change materials (PCMs) undergo cycling between two material states, such as solid-liquid, between two crystalline phases,  or crystalline-amorphous transitions, through the absorption and emission of thermal  energy \cite{ovshinsky1968reversible}. GeSbTe-based PCMs undergo reversible phase transitions between an amorphous state, which exhibits high electrical resistance, and a crystalline state, which exhibits low electrical resistance. The ability to switch between these high- and low-resistance states enables the storage and retrieval of binary information.  PCM materials have  applications in neuromorphic computing as non-volatile phase change random access memory (PCRAM) materials \cite{lee_ab_2008}. PCRAMs offer several advantages over traditional Si-based memory solutions including faster data read/write, good data retention, high thermal stability, and lower power consumption \cite{yamada_high_1987, Shportko2008}. PCRAMs based on chalcogenide materials are particularly advantageous due to their higher thermal stability, good cycling ability and fast phase transition speeds compared to other PCMs. Among the PCMs located along the pseudo-binary tie line between GeTe and Sb$_2$Te$_3$ \cite{yamada_high_1987}, Ge$_2$Sb$_2$Te$_5$ (GST) has emerged as a leading candidate for PCRAM applications, owing to favourable trade-offs between switching speed, data retention, stability, and low crystallisation temperature (150$^\circ$C)\cite{yamada_high_1987,xiong_low-power_2011,Lankhorst2005}. \\

For GST, the trigonal lattice (hexagonal crystalline group, \textit{P$\overline{3}$m1}) is the most stable crystalline polymorph, with a metastable rocksalt (FCC \textit{Fm$\overline{3}$m}) structure existing as an intermediate between the amorphous and hexagonal phases. Of the amorphous states for chalcogenide materials, GST has an intermediate ability to form the amorphous phase which lies between that of Sb$_2$Te$_3$, which has the highest tendency, and of GeTe, which displays the lowest ability for amorphization \cite{guo_review_2019}. Due to the lack of long-range atomic ordering, characterising the structure of the amorphous phase remains challenging. X-ray techniques, Raman spectroscopy, electron microscopy, and diffraction have been used to gain insight into the structure of the amorphous phase \cite{darrigo_mechanical_2018, frumar_atomic_2009, naito_local_2004, naito_direct_2010, bo_raman_2004}. It has been found that the amorphous phase has shorter bond lengths than that of the crystalline phase and despite these shorter bond lengths, the density of the amorphous phase is 5\% lower than that of the crystalline phase \cite{kolobov2004understanding}. 

For hexagonal GST, studies in the literature propose various stacking sequences, such as the Petrov Model, the Kooi de Hosson (KDH) model, the Ferro model, and the inverted Petrov model \cite{bala_recent_2023}. Two stacking models (Figure 1) are the most common: Te–Sb–Te–Ge–Te–Te–Ge–Te–Sb–Te (Petrov model) \cite{petrov1968electron} and Te-Ge-Te-Sb-Te-Te-Sb-Te-Ge (KDH Model) \cite{kooi_electron_2002}. Both of these models have nine atoms (2 Ge, 2 Sb, 5 Te) in the unit cell, stacked in layers along the \textit{c} axis.  Density functional theory (DFT) studies have reported the KDH model as the most stable by 2 - 6 meV/atom, albeit with a very small difference when considering the existence of vacancies and other point defects, where this small difference contributes to uncertainty over stacking sequences \cite{zhou_ab_2012, ZhouPhysRevLett.96.055507}.

 \begin{figure}
    \centering
    \includegraphics[width=1.0\linewidth]{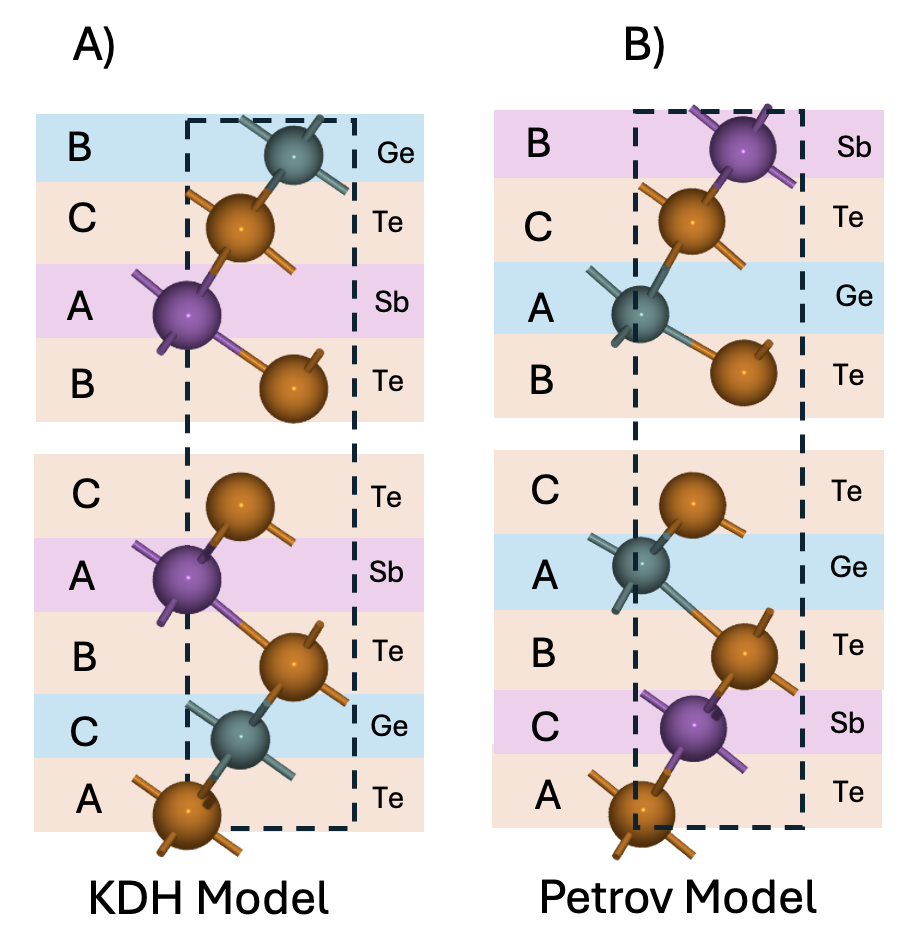} 
    \caption{Nine atom unit cells of KDH and Petrov structures of Ge$_2$Sb$_2$Te$_5$ (GST). Where A, B, and C denote different stacking positions with the unit cell and blue, purple, and orange spheres represent atoms of Ge, Sb, and Te, respectively.}
    \label{Figure1}
\end{figure}

\textit{Ab initio} computational studies have been used to investigate the properties of GST, but accurate description of the potential energy surface (PES) is needed. The harmonic approximation, where vibrations around the equilibrium bond distance are treated as a quadratic parabola, cannot accurately describe the PES for materials away from this equilibrium configuration. Consequently, it cannot accurately describe phenomena such as heat transport, thermal lattice expansion, and phase transitions, due to a lack of phonon-phonon interactions which are particularly relevant for GST materials \cite{lyeo_thermal_2006, duong_first-principles_2021, ghosh_significant_2023, deringer_2013_dft, AHN2020807}. Therefore, the ability to model these effects is needed for a thorough investigation of GST, thus methods beyond the limits of the harmonic approximation are needed. \cite{doi:10.1021/acs.jpcc.3c02863, ghosh_significant_2023, AHN2020807, 10.1063/5.0023476}. \\ 

Consideration of anharmonicity, that is deviations away from the harmonic approximation, is important as DFT studies have found that thermal transport in crystalline GST is dominated by low-frequency optical phonons with significant variation in harmonic and anharmonic thermal  transport properties \cite{ghosh_significant_2023}. Combined experimental and theoretical studies concluded that anharmonicity is closely related to resonant bonding and the local chemical structure, where anharmonicity was a factor in influencing phase-change properties \cite{AHN2020807}. Studies of anharmonicity in catalytic materials were also successful in quantifying the magnitude of anharmonicity contributions and used a metric, $\sigma$,  to resolve discrepancies between simultaneously and experimental vibrational spectra \cite{doi:10.1021/acs.jpcc.3c02863}. It is clear that quantification of anharmonic effects in GST is needed to rationalise properties such as phase-change behaviour and thermal transport, in addition for resolving different stacking models in GST. \\

Herein, we investigate the anharmonic behaviour of GST, whilst considering the structural properties, \textit{i.e.} the atomic stacking within the unit cell and van der Waals (vdW) interactions. Furthermore, we use the a metric for anharmonic quantification, $\sigma$, proposed by Knoop \textit{et al.} which is an evaluation of distortions on the local PES \cite{PhysRevMaterials.4.083809}. We propose that quantifying the magnitude of anharmonic effects would also provide insight into stability of the different stacking models proposed. To achieve this, we sample the PES of the Petrov and KDH models of GST at equilibrium, whilst considering the effects of different DFT exchange-correlation (XC). We then perform dynamical sampling of the PES through $ab$ $initio$ molecular dynamics simulations (AIMD), and from this we measure the strength of anharmonic effects for both Petrov and KDH models.   \\


\section{Methods}
\subsection{Quantification of Anharmonicity}

In this work, anharmonicity is defined as the deviation of a PES from the harmonic approximation which assumes the interatomic potentials take a quadratic shape (Figure 2). We examine anharmonicity in GST using the method proposed by Knoop \textit{et al.} \cite{PhysRevMaterials.4.083809}, which was adopted in previous work to study other materials \cite{doi:10.1021/acs.jpcc.3c02863}. The same workflow is employed here. It can be summarised as: i) evaluation of the harmonic PES through finite-difference phonon calculations ii) sampling a more realistic PES through \textit{ab initio} molecular dynamics (AIMD) simulations iii) through evaluation of force constants (FCs), quantify the deviation of the AIMD-PES with that of the harmonic approximation and obtain the standard deviation, $\sigma$. \\
 
 $\sigma$ is defined as the root-mean-squared error (RMSE) of the harmonic approximation in relation to the standard deviation of the force distribution: 

\begin{equation}
\label{Equation 1}
\sigma = \frac{\sigma[F^A]_t}{\sigma[F]_t} = \sqrt{\frac{\sum_{I,\alpha} \langle (F_{I,\alpha}^A)^2 \rangle_t}{\sum_{I,\alpha} \langle (F_{I,\alpha})^2 \rangle_t}}
\\
\end{equation}

where \( F \),  \( F^A\) are the total and anharmonic component $\alpha$ acting on atom \( I \) at time \( t \). A full description of the theoretical framework for the quantification of anharmonic is available in ref. \cite{PhysRevMaterials.4.083809}. \\

\begin{figure}
    \centering
    \includegraphics[width=0.9\linewidth]{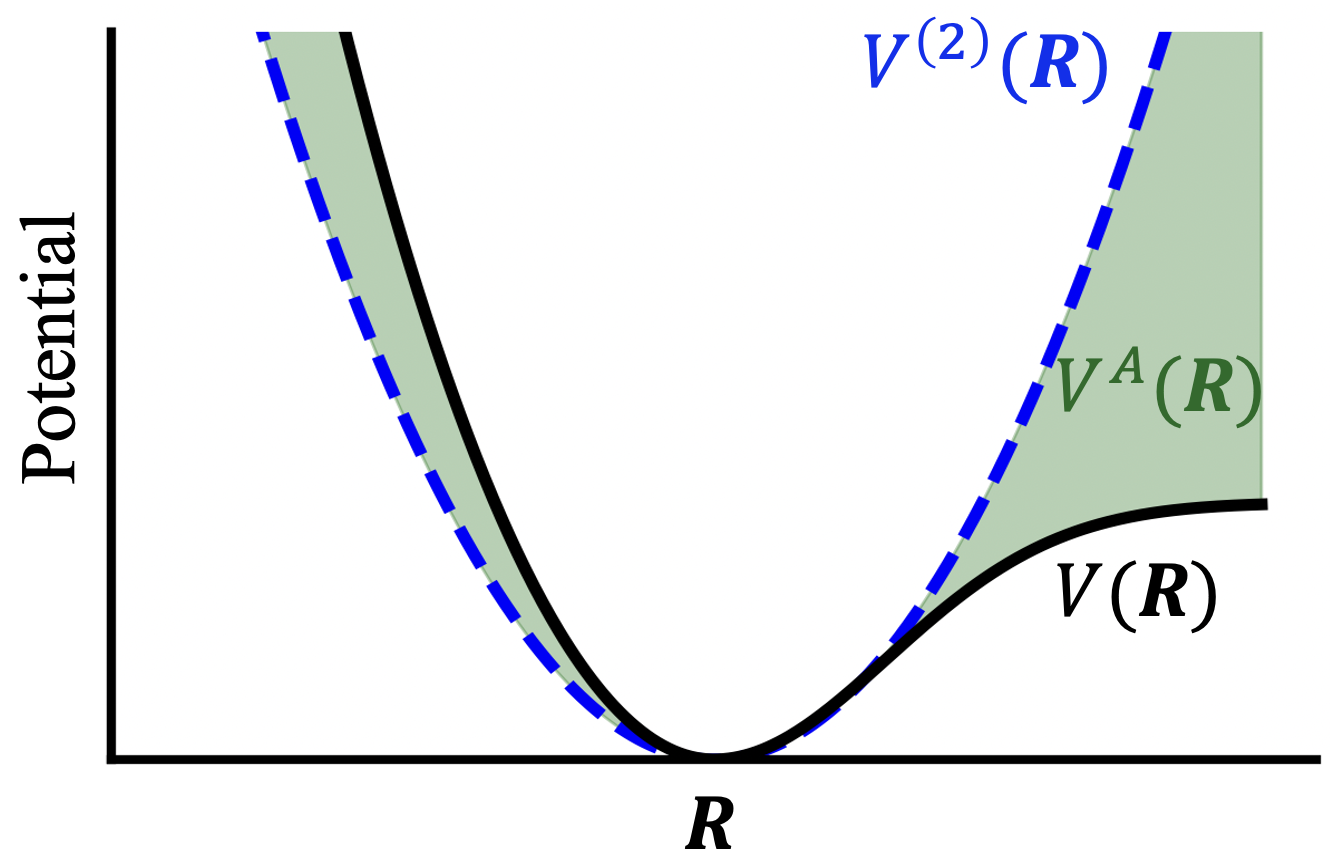} 
    \caption{Potential energy surface ($V(\mathbf{R})$), harmonic approximation ($V^{(2)}(\mathbf{R})$), and the deviation between the two, defined as anharmonicity ($V^{A}(\mathbf{R})$) shaded in green.}
    \label{Figure2}
\end{figure}

\subsection{Simulation Details} 

DFT calculations were performed with the Fritz Haber Institute Ab Initio Molecular Simulation (FHI-aims) software package \cite{blum_ab_2009}, which is an all-electron, full potential code. The general gradient approximation (GGA) exchange-correlation functional (XC) of Perdew-Burke-Ernzerhof (PBE), and the version re-parametrised for solids (PBEsol) were used with and without the Tkatchenko-Scheffler method (vdw(TS)) to account for dispersion corrections \cite{perdew_generalized_1996, perdew_restoring_2008, tkatchenko_accurate_2009}. Calculations were performed using a ‘light’ basis set of the 2010 release (Supporting Information (SI), Figure S1), where self-consistent field (SCF) cycle convergence reached when the sum of eigenvalues and change in charge density were below $10^{-6}$ $e/a_0^3$ and $10^{-6}$ eV, respectively. A converged Monkhorst-Pack k-point sampling grid of $4$x$4$x$2$ (SI, Figure S2) was used on unit cell structures of GST \cite{monkhorst_special_1976}. Calculations were performed spin-restricted and using the zeroth order regular approximations (ZORA) for relativistic treatments \cite{van_lenthe_relativistic_1994}. \\ 

Structures for the Petrov and KDH models (Table 1) were built and manipulated using the Atomic Simulation Environment (ASE) Python library \cite{hjorth_larsen_atomic_2017}. The hexagonal (HCP) GST unit cell was created using experimental lattice parameters for Petrov: $a,b = 4.20$ \AA, $c = 17.41$ \AA, and for KDH: $a,b = 4.25$ \AA, $c = 17.27$ \AA) where the crystallographic parameters for both models are shown in Table \ref{tab:models} \cite{petrov1968electron, kooi_electron_2002}.  All structure models were subject to full unit cell geometry and atomic optimisation, {\it i.e.}, lattice parameters and atom positions, using the Broyden-Fletcher-Goldfarb-Shanno (BFGS) algorithm with convergence reached when all the forces on all atoms were less than $0.01$ $eV \mathrm{\AA}^{-1}$ \cite{broyden_convergence_1970, fletcher_new_1970, goldfarb_family_1970, shanno_conditioning_1970}.\\

\begin{table}[ht]
\caption{\label{tab:models}Comparison of the crystallographic parameters of the KDH and Petrov models for GST. Both models share space group $P\bar{3}m1$ (164).}
\renewcommand{\arraystretch}{1.2}
\resizebox{\columnwidth}{!}{%
\begin{ruledtabular}
\begin{tabular}{lcc}
Model & Wyckoff Sites & Atomic Positions \\
\hline
KDH &
\begin{tabular}[c]{@{}c@{}}Te: 1a, 2d\\Ge: 2c\\Sb: 2d\end{tabular} &
\begin{tabular}[c]{@{}c@{}}Te(1a): (0,0,0); Te(2d): ($\tfrac{1}{3}$,$\tfrac{2}{3}$,$z$)\\
Ge(2c): (0,0,$z$); Sb(2d): ($\tfrac{1}{3}$,$\tfrac{2}{3}$,$z$)\end{tabular} \\
Petrov &
\begin{tabular}[c]{@{}c@{}}Te: 1a, 2d\\Ge: 2d\\Sb: 2c\end{tabular} &
\begin{tabular}[c]{@{}c@{}}Te(1a): (0,0,0); Te(2d): ($\tfrac{1}{3}$,$\tfrac{2}{3}$,$z$)\\
Sb(2c): (0,0,$z$); Ge(2d): ($\tfrac{1}{3}$,$\tfrac{2}{3}$,$z$)\end{tabular} \\
\end{tabular}
\end{ruledtabular}
} 
\end{table}

The anharmonicity metric, $\sigma$, for GST was calculated using FHI-vibes Python package. \cite{knoop_fhi-vibes_2020}. Harmonic force constants, $\Phi$, were obtained through the finite-difference methods \textit{i.e.,} through evaluation of phonons at the high-symmetry $\Gamma$ point, performed using the Phonopy Python package \cite{togo_first_2015}. For Phonon calculations, the Petrov and KDH unit cells were subject to further optimisations, with convergence reached when the forces on all atoms were less than $0.001$ $eV  \mathrm{\AA}^{-1}$ to eliminate residual forces. For full sampling of the PES, AIMD simulations were performed using the canonical (NVT) ensemble with ASE, using Langevin dynamics at 300 K. The anharmonicity metric, $\sigma$, was obtained from converged supercells of both Petrov and KDH models, 108 and 180 atoms, respectively, when performing phonon calculations, as negative phonon frequencies at the $\Gamma$-point disappear, which is where anharmonicity is measured (SI, Figure S3). Vibrational free energies  were calculated via Phonopy using phonon frequencies in the harmonic approximation at the $\Gamma$-point only. For the anharmonic PES, phonon vibrational frequencies were obtained from AIMD simulation trajectories via regression methods through hiPhive Python Package,\cite{eriksson_hiphive_2019} which presents a phonon representation of the anharmonic PES for thermodynamic analysis (Table S1). Spectroscopic properties were obtained via the vibrational density of states (VDOS) using a Fourier transform of the velocity autocorrelation function: 

\begin{equation}
\label{eqaution 2}
D(\omega) =
\int_{-\infty}^{\infty}
\frac{\nu(0)\cdot \nu(t)}{\nu(0)\cdot \nu(0)}(t)e^{-i\omega t}\,dt
\end{equation}

where $D(\omega)$ is the VDOS at frequency $\omega$ and $\nu$ is the velocity at time $t$. \\

Helmholtz free energy values were obtained for GST models via normal mode analysis from phonon and AIMD calculations, as defined by:

\begin{equation}
\label{equation 3}
F = \frac{1}{2} \sum_{qj} \hbar \omega_{qj} + k_B T \sum_{qj} \ln \left[ 1 - \exp\left( -\frac{\hbar \omega_{qj}}{k_B T} \right) \right]
\end{equation}

where $F$ is the Helmholtz free energy, $\sum_{qj}$ denotes the summation over all wavevectors $q$ in the Brillouin zone and all phonon branches \( j \), \( \hbar \) is the reduced Planck constant (\( \hbar = h / 2\pi \)), \( \omega_{qj} \) is the angular frequency of the phonon mode with wavevector \( q \) and branch \( j \), \( k_B \) is the Boltzmann constant, and \( T \) is the absolute temperature. The first term represents the zero-point energy of the system—energy that remains even at absolute zero due to quantum fluctuations. The second term accounts for the thermal excitation of phonons at finite temperature.

\section{Results and Discussion}

\subsection{Evaluation of the Potential Energy Surface}

For the evaluation of anharmonic effects, the choice of exchange–correlation (XC) functional and the inclusion of vdW interactions are important, especially for describing the Te–Te interactions in GST \cite{SINGH2018124, doi:10.1021/acs.inorgchem.7b01970, deringer_2013_dft, zhou_ab_2012, sa_first-principles_2014, cooley_first-principles_2020, duong_first-principles_2021, AIPAdv2018, IbarraHernPhysRevB.97.245205}. The results are highly sensitive to both the type and form of the XC functional in addition to the treatment of dispersion forces. This is typical for materials with weak vdW-type interactions \cite{PhysRevB.84.144117}, and therefore we have systematically compared different XC functionals at the GGA level and  dispersion corrections (Table \ref{tab:lattice}). 

\begin{table}[ht]
\caption{\label{tab:lattice}Lattice parameters (\AA) and Te–Te distances for the Petrov and KDH models of GST using various exchange--correlation (XC) functionals. Experimental values from Refs.~\cite{petrov1968electron,kooi_electron_2002} are included for comparison.}
\renewcommand{\arraystretch}{1.2}
\begin{ruledtabular}
\begin{tabular}{lcccc}
XC Functional & $a$ (\AA) & $b$ (\AA) & $c$ (\AA) & Te--Te (\AA) \\
\hline
\multicolumn{5}{c}{\textbf{Petrov Model}} \\
Exp.~\cite{petrov1968electron} & \textit{4.20} & \textit{4.20} & \textit{17.41} & \textit{3.60 - 3.75} \\
PBE            & 4.29 & 4.29 & 18.30 & 4.17 \\
PBEsol         & 4.23 & 4.24 & 17.20 & 3.66 \\
PBE+vdW(TS)    & 4.30 & 4.30 & 17.39 & 3.76 \\
PBEsol+vdW(TS) & 4.21 & 4.21 & 17.07 & 3.61 \\
\hline
\multicolumn{5}{c}{\textbf{KDH Model}} \\
Exp.~\cite{kooi_electron_2002} & \textit{4.25} & \textit{4.25} & \textit{17.27} & \textit{3.60 - 3.75} \\
PBE            & 4.33 & 4.33 & 17.84 & 4.32 \\
PBEsol         & 4.27 & 4.27 & 16.86 & 3.72 \\
PBE+vdW(TS)    & 4.31 & 4.31 & 17.40 & 4.04 \\
PBEsol+vdW(TS) & 4.25 & 4.25 & 16.69 & 3.62 \\
\end{tabular}
\end{ruledtabular}
\end{table}

As indicated in Table \ref{tab:lattice}, for both models, PBE slightly overestimates the lattice parameters in all directions. Relative to experiment, the in-plane lattice constants ($a,b$) are overestimated by about $+2\%$ for both KDH and Petrov models, while the $c$ axis is overestimated by approximately $+5\%$ for KDH and $+3\%$ for Petrov. In terms of the $c/a$ ratio, PBE gives values that are larger than experiment by around $+3\%$ (KDH) and $+1\%$ (Petrov).\\

For PBEsol, the $a,b$ lattice parameters are overestimated by roughly $+1\%$, whereas $c$ is slightly underestimated by about $-1\%$ for Petrov and $-3\%$ for KDH. Consequently, the $c/a$ ratio is underestimated by about $-5\%$ in the KDH model which is significant, while it is in excellent agreement (within $<1\%$) for the Petrov model. \\

The incorporation of vdW corrections (see PBE+vdW(TS) and PBEsol+vdW(TS) in Table \ref{tab:lattice}) improves the results for the lattice parameters. For PBE+vdW(TS), the $c$ axis contracts, reducing the overestimation and bringing the $c/a$ ratio for Petrov into closer agreement with experiment. For KDH, the addition of vdW corrections reduces both $a,b$ and $c$, improving the overall match. For PBEsol, vdW corrections further reduce the in-plane lattice constants, in much better agreement with experiment ($<0.5\%$ deviation), although $c$ remains slightly underestimated ($-1$ to $-2\%$). For the KDH model, PBEsol+vdW(TS) reproduces $a,b$ values within $<1\%$ of experiment, while $c$ is still underestimated by about $-3\%$. \\

Overall, all XC functionals reproduce the experimental structure within a few percent, where PBE tends to overestimate the $c$ parameter compared to PBEsol, which is closer to experiment. For the $a$ parameter, PBEsol produces results aligned with reported experimental values, and as with $c$, PBE overestimates the lattice constant. For PBE and PBEsol, the addition of vdW corrections contracts both $a$ and $c$ parameters, where PBEsol+vdW(TS) is most aligned with experiment for $a$.  This improvement of PBEsol over PBE is due to the reparameterization which reduces short-range repulsion and thereby contracts the lattice as atoms exhibit more attractive forces \cite{perdew_generalized_1996,perdew_restoring_2008}. \\

For accurate modelling of GST, it is important to capture the vdW-like Te-Te interaction  between the layers. Te, unlike lighter chalcogenide elements has a higher static polarizability, making induced dipoles more significant \cite{PhysRevA.110.042805}. This leads to secondary bonding interactions, where in Te the 5p $n$ (non-bonding) to $\sigma*$ antibonding orbital interaction is stronger \cite{C6CE00451B}. For PBE, in both the Petrov and KDH models, the Te-Te distance is greatly overestimated compared to experiment with 4.17  \AA, and 4.32 \AA, where reported experimental values are in the range of 3.60 \AA - 3.75 \AA  \cite{lee_ab_2008, sa_structural_2013, matsunaga_structures_2004}. The addition of vdW corrections contracts the Te-Te distance for both models, with Te-Te distance being reduced to $3.75$ \AA, and $4.04$ \AA for PBE+vdW(TS), where the former is in good agreement with experiment. Uncorrected PBEsol, gives Te-Te interactions of $3.66$ \AA\ and $3.72$ \AA\ for Petrov and KDH models, respectively, which are also within the reported experimental range. With PBEsol+vdW(TS), the Te-Te distance decreases to $3.61$ and $3.62$ \AA\ for Petrov and KDH models, which are also within the range of values reported in the literature \cite{lee_ab_2008, sa_structural_2013, matsunaga_structures_2004}. Overall, for both models PBEsol, PBE+vdW(TS) and PBEsol+vdW(TS) provide Te-Te distances commensurate with previous computational studies, where PBE alone overestimates Te-Te. \cite{Micoulaut2013,Bouzid2015, Schumacher2016, deringer_2013_dft, IbarraHernPhysRevB.97.245205, AIPAdv2018, Privitera2016} The poor performance of PBE is unsurprising and improvements for Te-Te distances seen in PBEsol is due to the aforementioned reparametrisation, where further contractions in Te-Te distances are seen when applying vdW(TS) corrections due to the inclusion of dispersion effects. PBEol consistently describes the structural parameters more accurately than PBE, and is possible going forward, since PES is ultimately highly dependent on the choice of XC functional. It is also seen that the inclusion of dispersion corrections, has a large effect on Te-Te, where accurate description of this interaction is needed, and may have consequences on anharmonic behaviour. Therefore, PBEsol with dispersion corrections (PBEsol+vdW(TS)) will be used to investigated the consequences of anharmonicity on the Petrov and KDH stacking models of GST.

\subsection{Anharmonic Effects in Ge$_2$Sb$_2$Te$_5$}

With the overall anharmonicity metric, $\sigma^A$ (Equation 1), \textit{i.e.} for the whole material, the Petrov model has an anharmonicity score of  $0.41$, meaning that anharmonicity contributed to $41$\% of the forces (Figure 3A). For the KDH model, a less pronounced anharmonicity is seen with $\sigma^A$ value of $0.28$. This means that the KDH model is only moderately anharmonic with $\sim$ $30$ \% of forces containing anharmonic contribution. The difference between anharmonicity between Petrov and KDH models is due to the stacking sequence. It is reported that the KDH model inherently better accounts for vdW interactions owing to the stacking sequence, \cite{https://doi.org/10.1002/adfm.201803380} which has stacking  of Te-Ge-Te-Sb-Te either side of the vdW-gap .  In the Petrov model, the stacking around the vdW-gap is Te-Sb-Te-Ge-Te and with just under half the forces not being accurately represented by the harmonic model, it is concluded that this stacking sequence is more unstable by pulling the PES away from an ideal parabolic shape. \\ 



\begin{figure}[h!]
    \centering
    \includegraphics[width=1.0\linewidth]{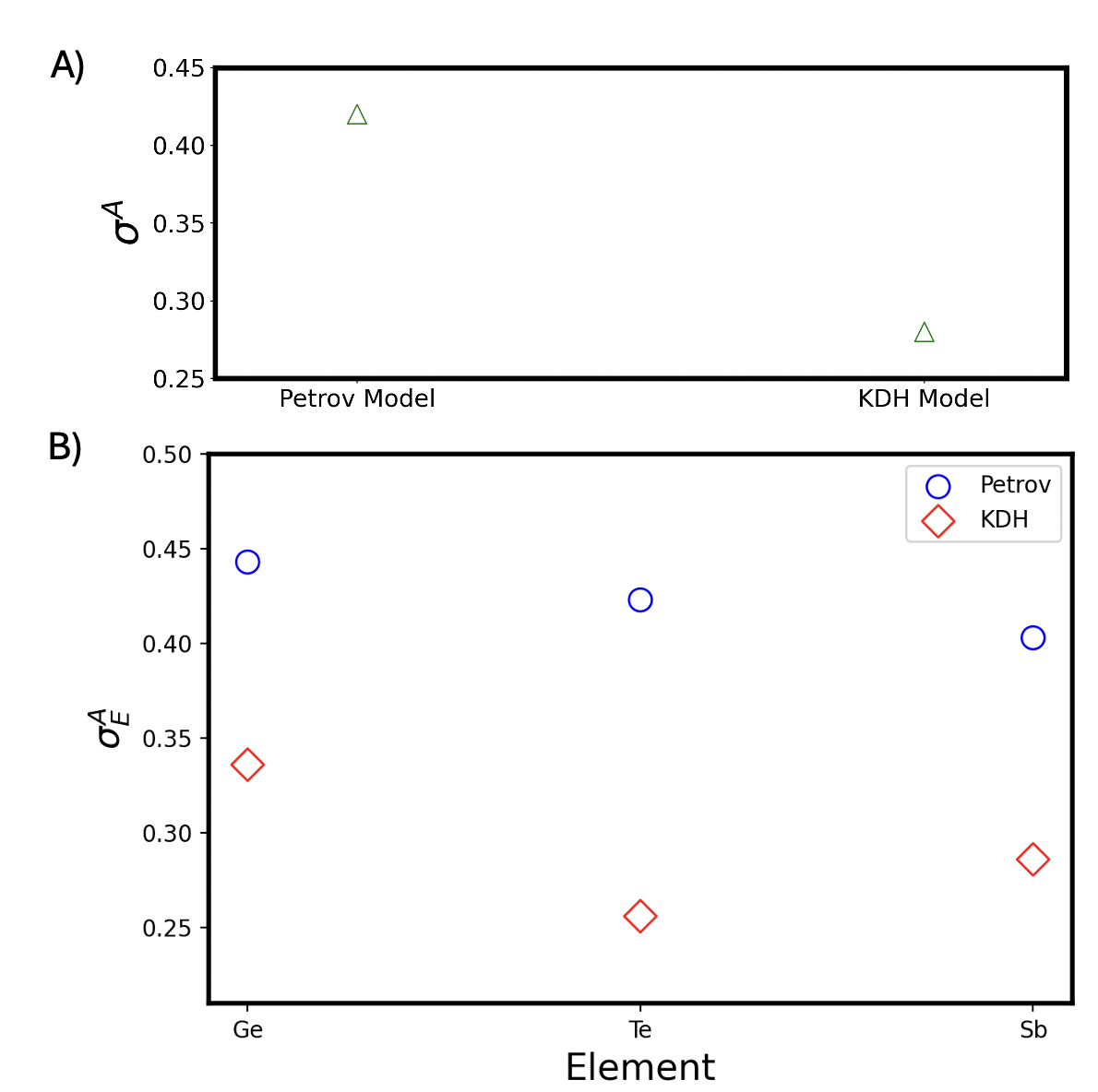} 
    \caption{A) Total anharmonicity, $\sigma^A$, for Petrov and KDH structure models. B) ELement resolved anharmonicity, $\sigma^A_E$, for Petrov (blue circle) and KDH (red diamonds) models. }
    \label{Figure3}
\end{figure}

 The element-resolved anharmonicity, $\sigma^A_E$ (Figure 3B) measures the anharmonic contributions per element in the material. For the Petrov model, all elements present within the material display moderate anharmonicity ($> 0.4$); however, Ge displays the greatest anharmonicity, with an order of $Ge > Te > Sb$ with values of $0.44$, $0.42$, and $0.40$, respectively. For the KDH model, Ge again is the most anharmonic element with $\sigma^A_E$ values of $0.35$, where Sb has a greater $\sigma^A_E$ than Te, with values of $0.29$ and $0.26$ for Te and Sb, respectively, bringing the total order of anharmonicity to $Ge > Sb > Te$. The difference in trend between Petrov and KDH models supports findings that the stacking sequences affects the PES and subsequently anharmonic behaviour. As the Petrov model has Ge atoms directly connected to Te around the vdW-gap this suggests that being directly connected to the vdW-gap increases anharmonicity. As for the KDH model, where Sb is connected to the vdW-gap, $\sigma^A_E$ for Sb is increased relative to Te, where in the Petrov model the trend is reversed. Although for both models Ge remains the most anharmonic element, in the KDH model the anharmonicity of Ge is much lower than that of the Petrov, which may be be due to lack of bonding to the Te atoms of the vdW-gap. Furthermore, the overall higher anharmonicity of Ge could be a consequence of smaller atomic mass, where larger vibrational amplitudes of atomic displacements distort the local PES away from an ideal parabolic shape. \\

For vibrational mode-resolved anharmonicity, $\sigma^A_s$, (Figure 4), shows that anharmonicity is dispersed from in the range of $\sim$ $0.2$ to $0.4$ for vibrational frequencies between 3 to 6 THz, meaning these forces are moderately anharmonic across all models. Below $3$ THz anharmoncity is increased slightly, especially for the KDH model, where frequencies with $\sigma^A_s$ of $0.6$ and $0.8$ are seen which are notably absent in the Petrov model. Furthermore, for both models a low-frequency single vibrational mode, corresponding to acoustic phonons, has an anharmonicity score of $1$ $i.e.,$ $100$\%, meaning that this mode is dominated by anharmonic behaviour. The higher anharmonicity of low-frequency acoustic phonons is intuitive, as heat transport mechanisms, which are not described in the harmonic approximation, are dominated by acoustic phonons \cite{ghosh_significant_2023}. This is observed in Figure 4, as the distribution of $\sigma^A_s$ decreases slightly with increasing vibrational frequency. However, $\sigma^A_s$ remains moderately high at higher vibrational frequencies (optical phonons), with the anharmonicity of vibrational modes remaining between $\sigma^A_s$  $0.2$ and $0.4$. Recent studies in the literature have found that hexagonal GST unusually shows a high degree of optical phonon heat transport \cite{ghosh_significant_2023, Mukhopadhyay2016}. The high $\sigma^A_s$ values seen in Figure 4 support these findings as high frequency optical phonons remain strongly anharmonic, although key differences are seen between both models. It is seen that the KDH modes are more anharmonic between 1-3 THz compared to the Petrov model, where the for the Petrov model, the modes are more anharmonic compared to KDH in the range of 4-5 THz. The results confirm that atomic ordering does affect anharmonic behaviour, particularly in relation to the vibrational modes. Indeed, phonon band structures showing atomic contribution for each stacking model (SI, Figures S4, S5) show in the Petrov model optical bands are dominated by Ge, where in the KDH model these high frequency modes are predominately populated by Sb contributions, in agreement with other studies \cite{ghosh_significant_2023} and highlight the variations in vibrational behaviour observed when changing the stacking order in between the Te layers. Nevertheless, better insight into the relationship between the stacking sequences is needed, therefore we consider the affects of the local chemical structure on anharmonicity. \\

 \begin{figure}[h!]
    \centering
    \includegraphics[width=1.0\linewidth]{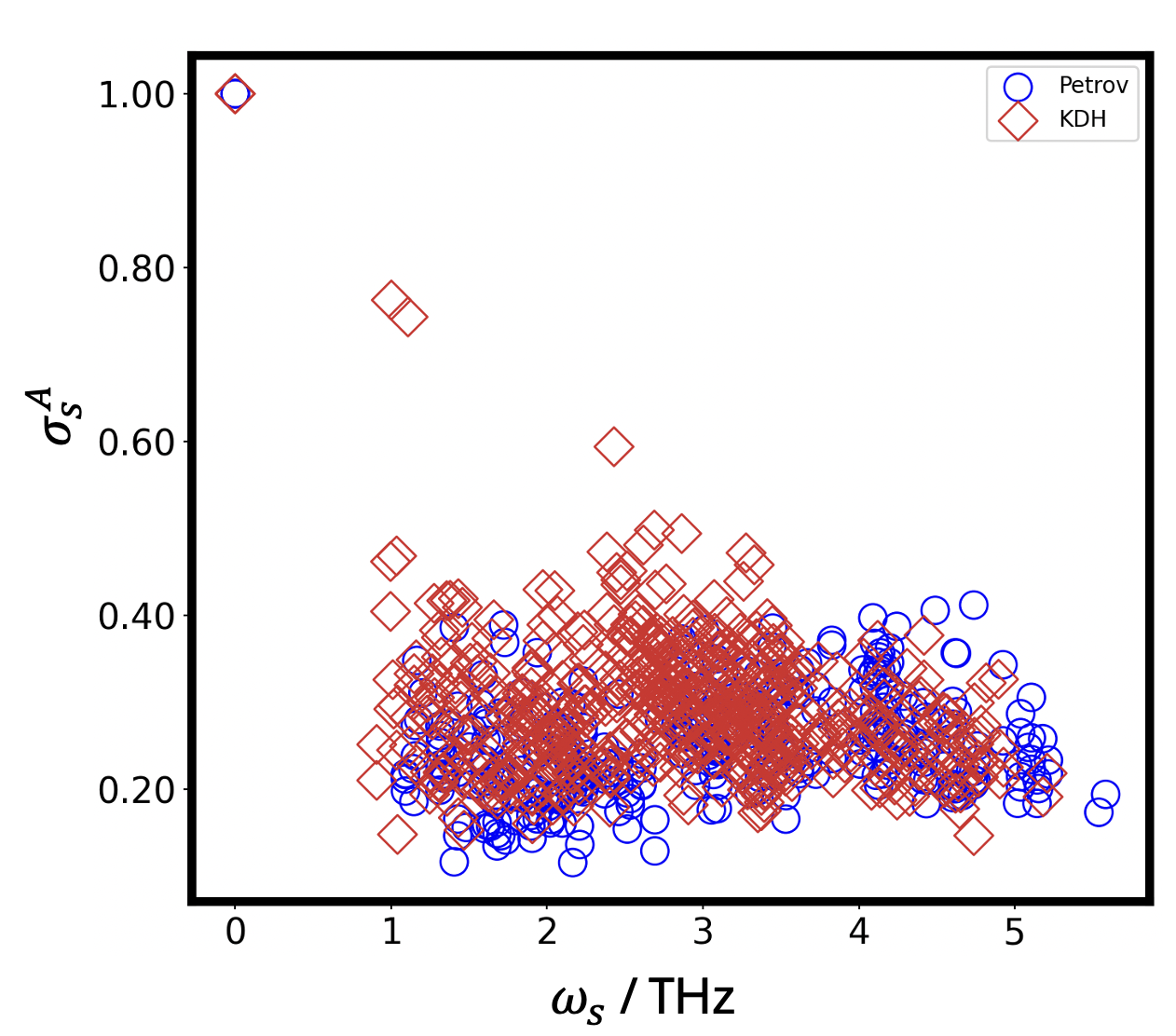} 
    \caption{Anharmonicity resolved by vibrational modes, $\sigma^A_s$. }
    \label{Figure6}
\end{figure}

By spatially resolving  anharmonicity on the simulation cell, the per-atom anharmonicity is obtained, which can provide insight on localised anharmonic effects,  as shown in Figure 5. In the case of the Petrov model (Figure 5, top), Sb atoms on the top of the cell have high anharmonic character, where neighbouring Te atoms in the layer below, also have comparable high degree of anharmonicity at $\sim 0.44$. This is also true for some Te atoms in the layer below, indicating that atoms connected to Te are anharmonic. The lowest anharmonicity is seen with bottom Te atoms in the ninth layer which display anharmonicity of $\sim 0.20$. With the KDH model,  anharmonicity is lower, with Ge atom layer being consistently more anharmonic, albeit moderately. Interestingly, atoms away from the Te-Te gap also display high anharmonicity, indicating that long range effects on the atomic bonding away from the gap is observed. This corroborates the results from Figure 3, where the Petrov model has a higher overall anharmonicity score compared to the KDH model,  and that element resolved anharmonicity shows Ge being the most anharmonic element for the KDH model. \\

\begin{figure}[h!]
    \centering
    \includegraphics[width=1.0\linewidth]{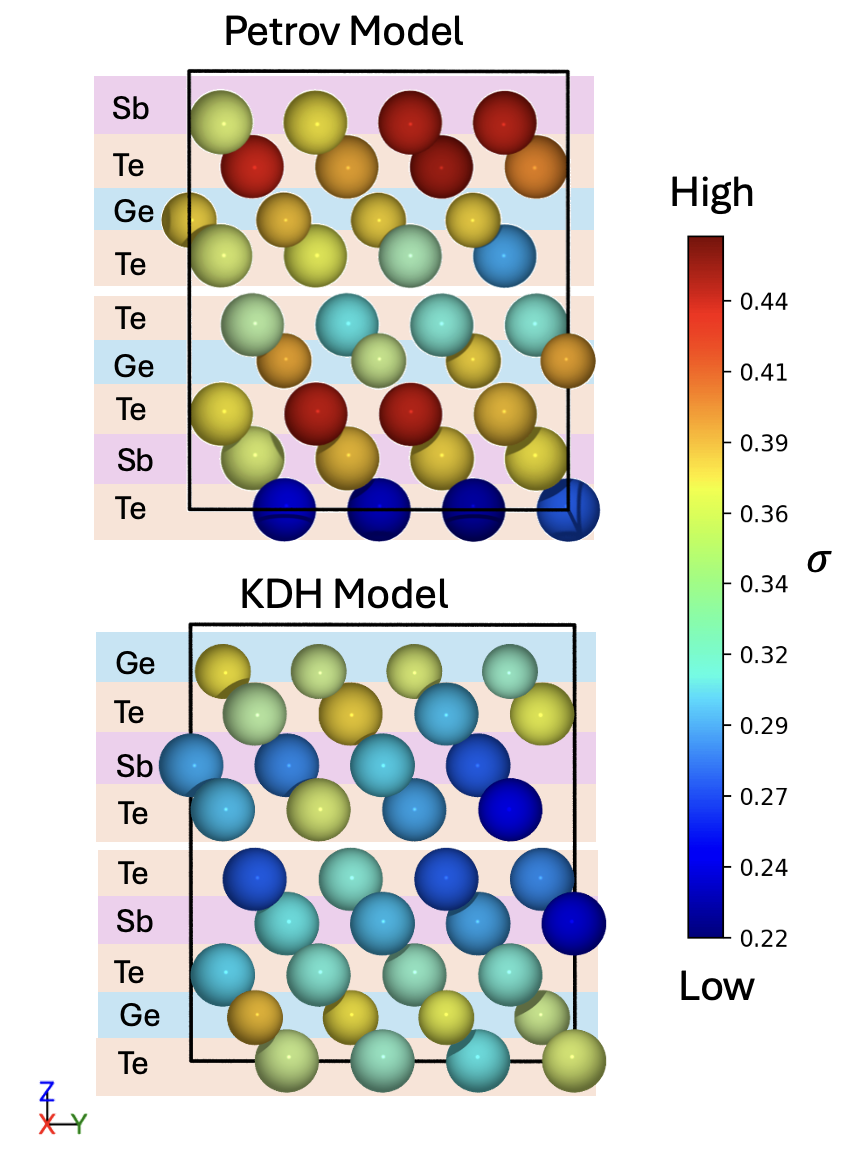} 
    \caption{Color plots of atom resolved anharmoncity for Petrov (top) and KDH models (bottom) for simulation cells ($108$ atoms) along the $ZY$ plane. A reduced simulation cell is presented for KDH model for the purpose of visualisation.}
    \label{Figure5}
\end{figure}

These results indicate that Te-Te interactions are the main driving force of anharmonic behaviour observed due to the weaker bonding. Analysis of the distributions of atomic displacements from AIMD simulations (Figure 6), indicate that anharmonicity is strongly directional. For both stacking models, Te atoms displays greater displacements, $|u|$, in the Z-direction, indicating enhanced soft vibrational modes arising from the interlayer Te-Te interactions of the vdW-gap.  For the Petrov model, in the X-direction, Sb displacements are  greater than in the case of KDH, however, the greater displacements are seen in the Z-direction. Considering that $\sigma^A_E$ shows a pronounced anharmonicity for Ge atoms, the type of bonding characteristics and subsequent displacements play a key role and strongly indicate that the Petrov stacking sequence (Te-Sb-Te-Ge-Te-Te-Ge-Te-Sb) is more unstable than KDH stacking (Te-Ge-Te-Sb-Te-Te-Sb-Te-Ge). In addition to the vdW-gap itself, the atom connected to the Te-Te vdW gap may influence anharmoncity. \\

Analysis of trajectories from AIMD simulations of the Petrov model (SI, Figure S7) show that the Te-Te distances for PBEsol remain higher than PBEsol+vdW(TS) throughout the simulation, with average distances of $3.64$ \AA,  and $3.60$ \AA, respectively, \textit{c.f.} Table \ref{tab:lattice} where reported values of $3.66$ and $3.61$ \AA are seen in optimized structures,  respectively.  Furthermore, it is seen that in general PBEsol+vdW(TS) contains the larger oscillations of bond distances in the MD simulations, compared with PBEsol. Therefore, vdW corrections for the Petrov model perturbs the PES away from a ideal parabolic potential \textit{i.e.,} the harmonic approximation, resulting in the increase of anharmonic behaviour. This increased oscillation due to inclusion of attractive and repulsive forces, distorts the bonding with neighbouring Ge atoms around the vdW gap, which in turns increases the anharmonicity as seen with $\sigma^A_E$, where the atomic weight of Ge may influence susceptibility to anharmonic displacements.

\begin{figure}[h!]
    \centering
    \includegraphics[width=1.0\linewidth]{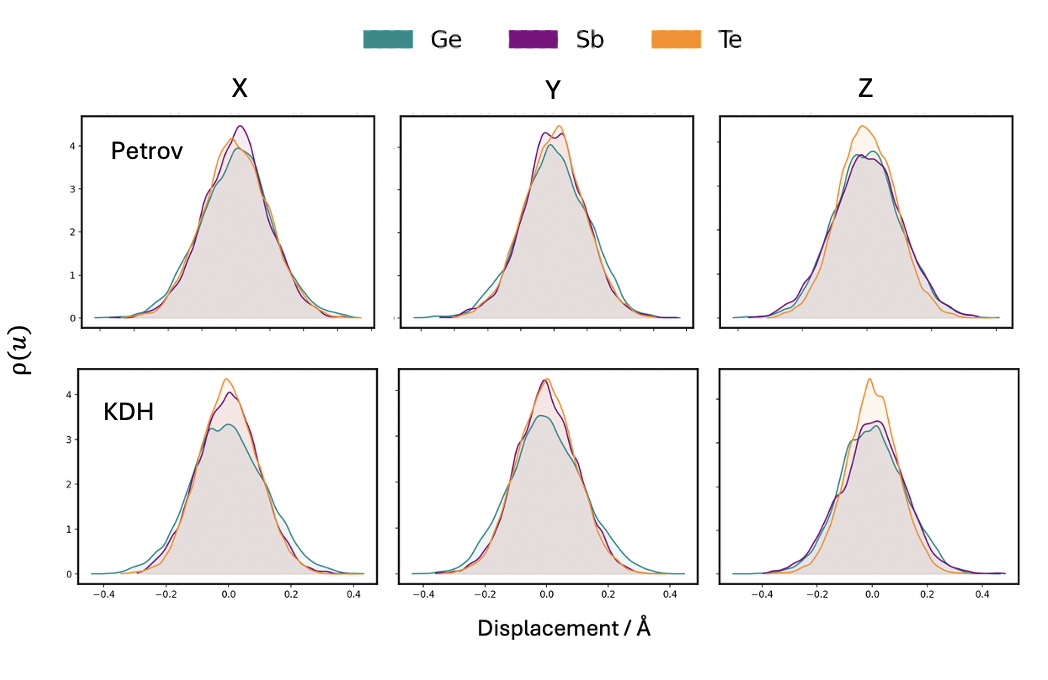} 
    \caption{Distribution of displacement $\rho(u)$ observed during AIMD simulation at 300 K.}
    \label{Figure8}
\end{figure}

The rationalize the varying strength of anharmonic effects displayed with the Petrov and KDH modes analysis of the electronic structure displayed by each model is presented. Analysis of the electronic density of states (DOS) (Figure 8), shows little difference between the total DOS for the stacking models of GST. Both Petrov and KDH models display similar electronic features in the DOS, namely, broad and dispersed electronic states around the Fermi level, where states are dominated by Te atomic orbitals. Nevertheless, some differing features are observed when looking at orbital contribution for each element; in particular, the Petrov model displays discrete bands in the region of -10 to -6 eV, where the equivalent region is more dispersed in the KDH model. Furthermore, for the KDH model increased orbital mixing is observed, with hybridization between the Te p-orbital and the Sb p-orbital, with an increased contribution from the Ge-p obital in the region just above the Fermi level.  This suggests that there is weaker interactions and overlap between orbitals in this region in Petrov model compared to the KDH. Similarly, discrete peaks are seen in the DOS above the Fermi level in the region of 0 to 2 eV, which is also attributed to weaker bonding. \\

\begin{figure}[h!]
    \centering
    \includegraphics[width=1.0\linewidth]{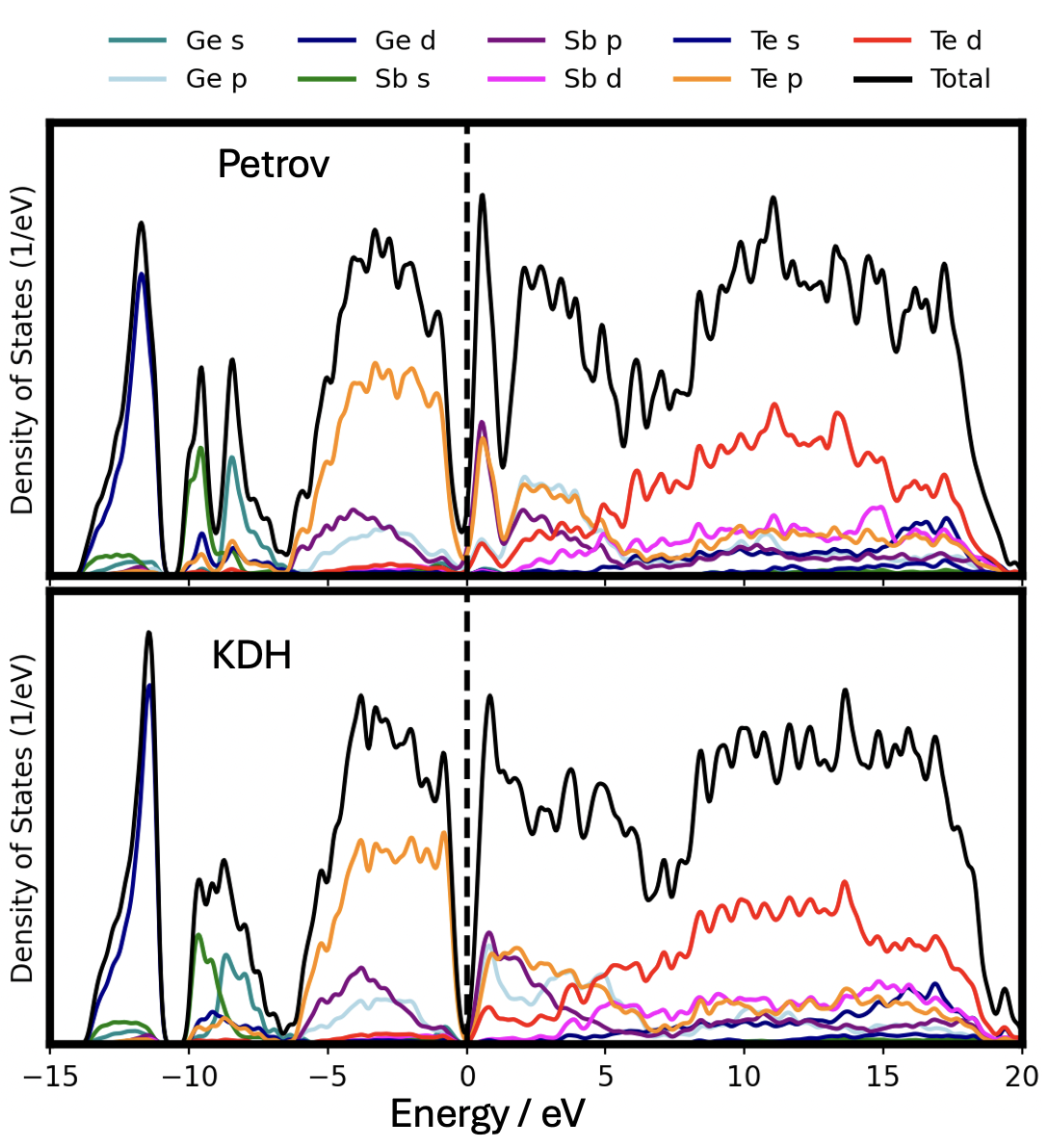} 
    \caption{Electronic density of states (DOS) plots for Petrov and KDH models with and without vdW(TS) corrections.}
    \label{Figure9}
\end{figure}


 Analysis of the electron density (Figure 7) provides more detail and indicates that the changes in the electronic structure around the Te-Te gap influences the anharmonic behaviour observed. For the KDH model, larger electron density is seen on the lower Te of the vdW gap compared with the Petrov model and weaker overlap between the gap-Te and the bottom Ge atom is observed, meaning more electron density from the Te is present within the inter Te-Te space. For Petrov model, very dispersed electron density is seen between the lower Te of the vdW gap and bottom Sb atom. Where the electron density on the lower  Te atom od the vdW-gap is lower compared to the KDH model. Furthermore, the movement of electron density away from the bottom Sb layer could be the driving factor of increased anharmoncity, which is seen in Figure 6, as $\sigma$ is strongly concentrated on the neighbouring atoms around the Te-Te gap of the Petrov model with PBEsol+vdW(TS). This further suggests that Petrov stacking (Te-Sb-Te-Ge-Te-Te-Ge-Te-Sb) is more unstable than KDH stacking (Te-Ge-Te-Sb-Te-Te-Sb-Te-Ge), as addition of dispersion correction for the Petrov model significantly change the electronic structure and subsequently the PES and strength of anharmonic behaviour. \\

\begin{figure}[h!]
    \centering
    \includegraphics[width=1.0\linewidth]{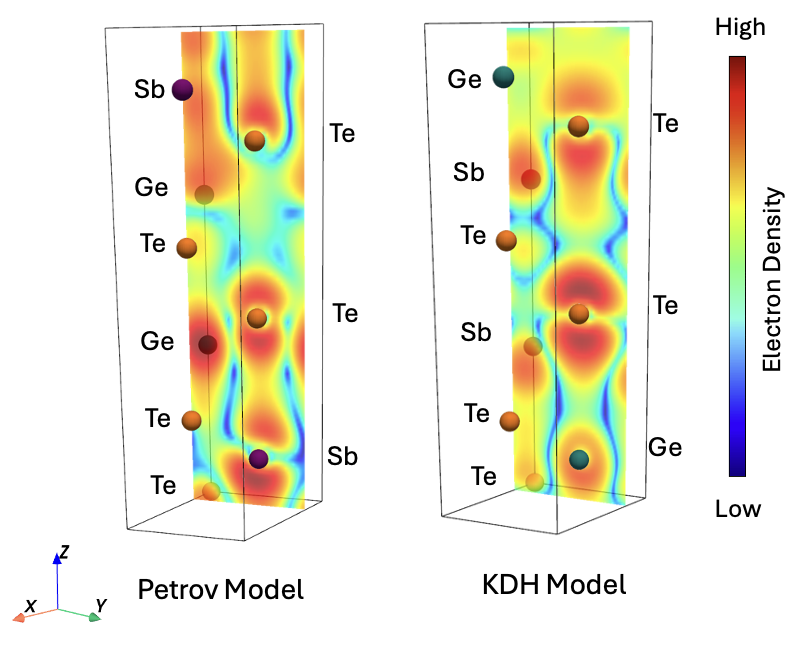} 
    \caption{Electronic density plots for Petrov and KDH models with and without vdW(TS) corrections.}
    \label{Figure10}    
\end{figure}

\subsection{Anharmonic Effects on Thermodynamic and Spectroscopic Properties}

The difference in vibrational properties from harmonic (phonon) and anharmonic (AIMD) methodologies directly impact the assessment of thermodynamic properties. The results of the free energy of both Petrov and KDH models in a harmonic and anharmonic frameworks at $300$ K are noted in Table \ref{tab:energies} and calculated according to Equation \ref{eqaution 2}. \\

\begin{table*}[ht]
\caption{\label{tab:energies}Comparison of harmonic and anharmonic free energies (per atom) for the KDH and Petrov models.}
\renewcommand{\arraystretch}{1.2}
\begin{ruledtabular}
\begin{tabular}{lcc}
Stacking Model &  Harmonic (eV/ atom [kJ mol$^{-1}$ / atom]) & Anharmonic (eV/ atom [kJ mol$^{-1}$/ atom]) \\
\hline
KDH     & $-0.0653$ [$-6.30$] & $-0.0682$ [$-6.58$] \\
Petrov  & $-0.0633$ [$-6.11$] & $-0.0674$ [$-6.50$] \\

\end{tabular}
\end{ruledtabular}
\end{table*}

In Table \ref{tab:energies}, it is seen that harmonic approximation consistently underestimates the free energy of GST compared to an anharmonic framework. This indicates that when considering the thermodynamic stability of the difference stacking structures, accounting for anharmonic effects is important. For both harmonic and anharmonic framework, the free energy of the KDH model is lower, suggesting that this is the most stable stacking sequence, which is commensurate with previous results in this work and studies in the literature, which find that the KDH model is energetically more stable (2-6 meV) \cite{zhou_ab_2012, ZhouPhysRevLett.96.055507}. In this work, considering the Helmholtz free energy, as opposed to the potential energy, we find the KDH model is an order of magnitude more stable than studies in the literature (2-6 meV) with the KDH model being 88 meV (anharmonic) and 65 meV (harmonc) more stable than the Petrov model. From these results it is clear that the harmonic model fails to capture important thermodynamic properties leading to the underestimation of free energy, because of the omission of the temperature dependence of vibrations frequencies. Since the harmonic approximation fails to accurately describe the temperature dependence of equilibrium properties, it cannot account for the thermal lattice expansion of materials. Therefore to assess the thermal lattice expansion of both Petrov and KDH models of GST, there is a need to move beyond the limitation of the harmonic model and herein we employ to AIMD simulations, which account for anahrmonic effects, to assess the thermal lattice expansion. \\

From AIMD simulations, drastically different thermal lattice expansion (Figure 9) is observed for both Petrov and KDH model with PBEsol. The lattice volume was calculated at temperatures at 20 and 300-800 K from AIMD simulations. For the Petrov model, an increase of $\sim 8.00$ $\mathrm{\AA}^3$ is observed from $20$ to $800$ K from $265.00$ to $272.84$ $\mathrm{\AA}^3$. For the KDH model a lattice expansion of $\sim 10.00$ is seen, from $261.75$ to $272.22$ $\mathrm{\AA}^3$. Interestingly, although the KDH model undergoes a greater expansion compared to the Petrov model, at $800$ K both models have lattice volumes of $\sim 272$ $\mathrm{\AA}^3$, suggesting that the Te-Te layer suppresses the expansion mechanism, especially at higher temperatures due to low-energy phonon modes associated with vdW interactions \cite{sklenard_electronic_2021}. Furthermore, nearly identical volume expansion is observed for the Petrov model using the quasi-harmonic approximation (QHA), Figure S8, however, for the KDH model the QHA predicts slight lattice contraction, contradictory to results from AIMD simulations. This suggests that the harmonic and quasi-harmonic approximations fail to capture the complexities arising from vdW interactions in GST, and that full considerations of anharmonicity is needed for modeling this material. \\

\begin{figure}[h!]
    \centering
    \includegraphics[width=1.0\linewidth]{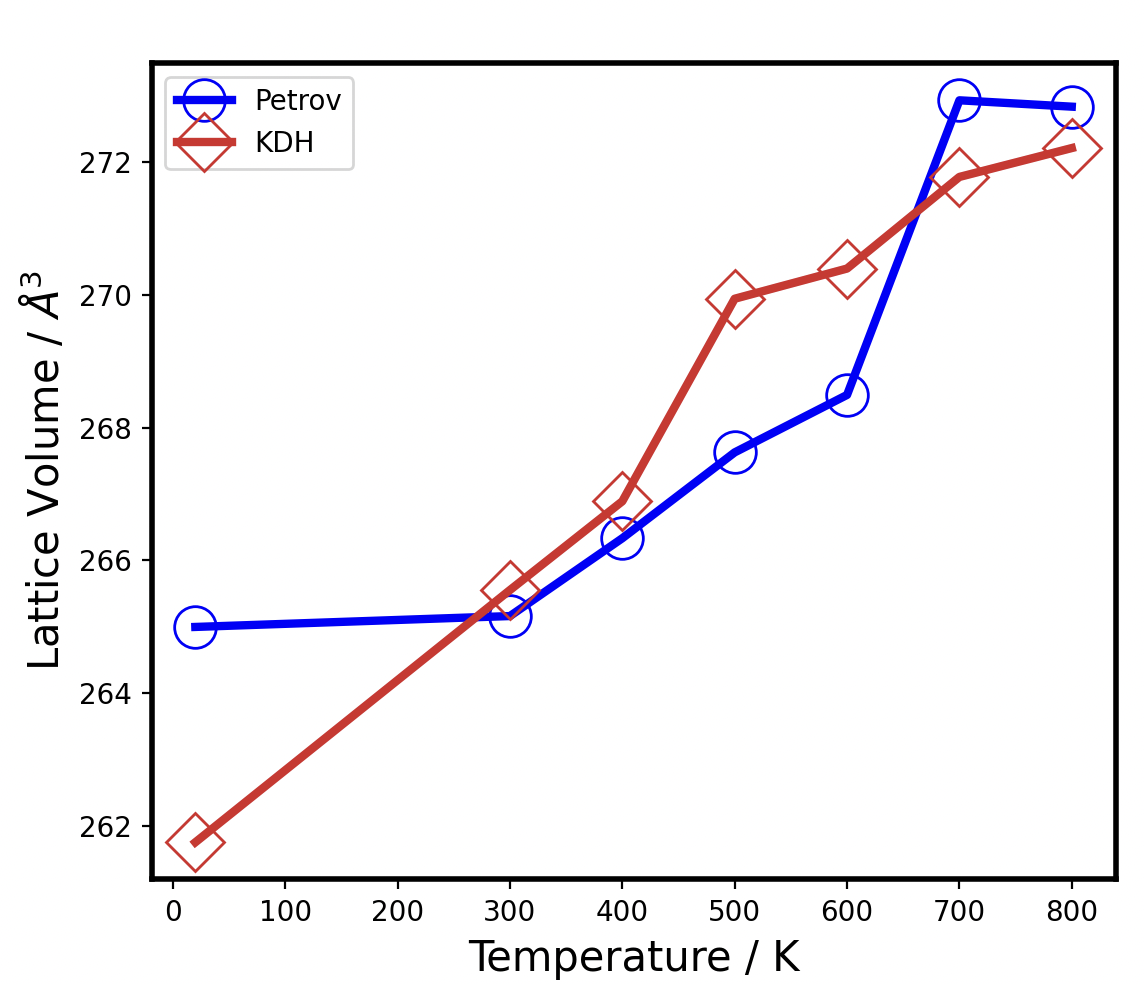} 
    \caption{Thermal lattice expansion 20-800 K of Petrov and KDH model with PBEsol and PBEsol+vdW(TS) from AIMD simulations.}
    \label{Figure11}    
\end{figure}


Finally, we compare the vibrational DOS obtained from calculations with experimental studies in the literature. We find that incorporation of anharmonicity better replicates vibrational spectra from experiment compared to the harmonic approximation. Figure 11 shows the vibrational density of states (VDOS) obtained from AIMD simulations which account for anharmonic effects and the VDOS from phonon calculations using only the harmonic approximation. Both methods replicate the key vibrational features present in infrared spectra from experiment \cite{bo_raman_2004, yin_microstructure_2018, andrikopoulos_raman_2007, wei_grayscale_2017, vinod_structural_2015, kozyukhin_structural_2013, bala_ga_2022, xu_optical_2018, talochkin_optical_2021, mio_role_2017, abou_el_kheir_unraveling_2024}, namely the main peaks at $100-120$ $\mathrm{cm^{-1}}$ and $\sim 150$ $\mathrm{cm^{-1}}$, which are characteristic of GST. For both models, the phonon modes around $\sim 100$ $\mathrm{cm^{-1}}$ are due to $\mathrm{E}$-type vibrations along the \textit{ab} plane and phonon modes at around beyond $150$ $\mathrm{cm^{-1}}$ are due $\mathrm{A}$-type vibrations along the \textit{c} plane. However, differences between the stacking models are observed, with an intense peak at $\sim 100$ $\mathrm{cm^{-1}}$ present in the KDH model which is significantly lower in the Petrov stacking. In the KDH model, the intense peak at $106$ $\mathrm{cm^{-1}}$ is due to the vibrational modes of the bottom Te-Ge-Te sublattice, which is absent in the Petrov stacking model. Furthermore, additional peaks are in seen in the Petrov model at $\sim 150$ $\mathrm{cm^{-1}}$ is due to the $\mathrm{A}$-type vibrations to the bottom the bottom Te-Sb-Te sublattice. Phonons below $\sim 100$ $\mathrm{cm^{-1}}$ are primarily from soft vibrational modes of the lattice.  The types of vibrational modes observed and differences between stacking orders is commensurate with other computational investigations \cite{sosso_vibrational_2009, Mukhopadhyay2016}. We note, however, that from the anharmonic VDOS, the Petrov model between replicates the vibrational positions seen from Raman spectra, but the KDH model better reflects the difference in peak intensities, indicating that in experiment a combination of both stacking sequences could be present, which is possible given the relatively small difference in free energy between both models of $88$ meV as calculated with the anharmonic approximation. The similarity between the anharmonic VDOS and Raman spectra supports that an anharmonic treatment of GST is crucial when bridging theoretical and experimental studies.

\begin{figure}[h!]
    \centering
    \includegraphics[width=0.9\linewidth]{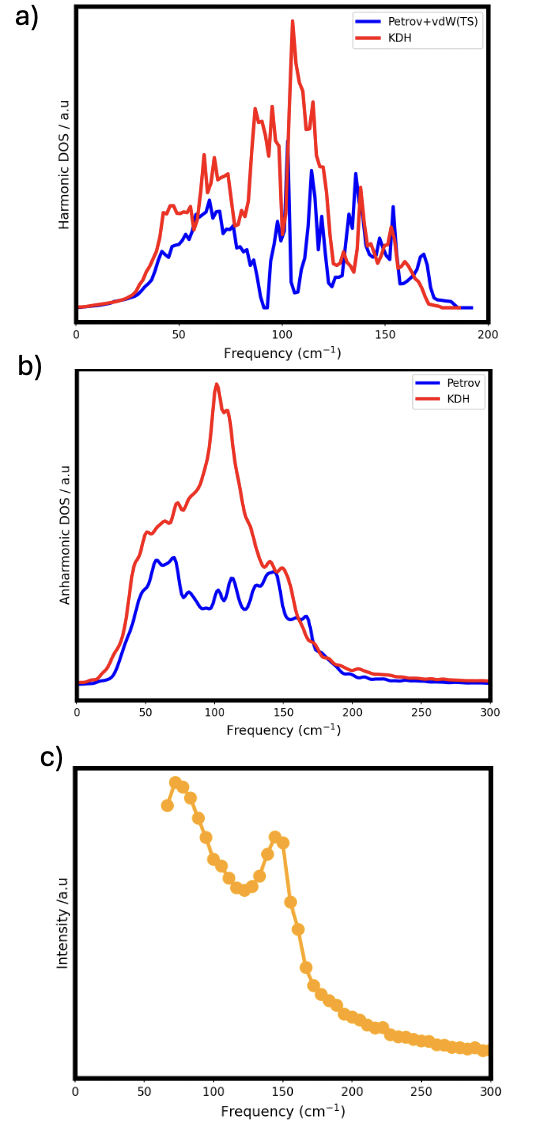} 
    \caption{a) harmonic vibrational density of states (VDOS) from finite difference method b)anahrmonic vibrational density of states (VDOS) from AIMD simulations c) Raman spectra of GST from experiment reproduced from Ref[65].}
    \label{Figure12}    
\end{figure}

\section{Summary and Conclusion}

In this work, an investigation of the effects of global and local anharmonicity on different stacking models, Petrov and KDH, for GST was undertaken. By comparing the harmonic approximation with the PES obtained from MD simulations at 300 K, a score for anharmonicity, $\sigma$, was derived which was used to quantify anahrmonic contributions in Petrov and KDH stacking models. Overall, the Petrov model  displayed the highest anharmonicity, with $\sigma$ of 0.41, indicating that under half the forces have anharmonic contribution. For the KDH model lower anharmonicity is found with essentially with $\sigma$ of $0.29$. For element- resolved anharmonicity, $\sigma^A_E$, Ge is the most anharmonic element for both models, where in the Petrov model the ordering of elements is $Ge > Te > Sb$. For the KDH model an ordering of $Ge > Sb > Te$ is observed. Mode-resolved anharmonicity, $\sigma^A_S$, which gives insight into anharmonicty per vibrational mode, where it is observed that high frequency optical phonons remain significantly anharmonic suggesting these may affect in thermal transport mechanisms. Moreover, atom-resolved anharmonicity indicates that anahrmonicity is localised in nature rather than uniform across the structure, and in the case of Petrov model is concentrated around the vdW-gap. This is corroborated by examination of the electron density, where for Petrov model is concentrated below the vdW-gap strengthening the interaction of atoms within the layers, as opposed to the KDH model, where greater electron density is seen in between the Te-Te atoms of the vdW-gap. Together, these results indicate that KDH stacking (Te-Ge-Te-Sb-Te-Te-Sb-Te-Ge) is more stable than Petrov stacking (Te-Sb-Te-Ge-Te-Te-Ge-Te-Sb).  \\

By examining the free energy confirms that the KDH model is the more thermodynamically stable stacking sequence and that the harmonic approximation underestimates the free energy compared to an anharmonic framework.  Finally, by comparing the harmonic and anharmonic VDOS with infrared spectra from experiment, we show that although both methods reproduce key vibrational features from experiment, accounting for anharmonic effects provides results in better agreement with experiment.  Our results show that anharmoncity plays varying roles in different stacking models of GST, therefore the choice of model must be carefully considered when investigating properties such as thermal transport in GST. Furthermore, we  note that due to limitations of computational methods, particularly the high computational overhead of AIMD simulations, our investigation was restricted to the semi-local GGA level, which are known to produce shallow PES and red-shifted vibrational modes. With the continuing improvement of computational methods,  specifically machine learning interatomic potentials (MLIPS), further investigation will be needed to assess the effects of anharmonic behaviour at higher levels of DFT, as ultimately the choice of theory dictates the PES.

\section{Associated Content}

The accompanying supporting information contains further details of calculation methods. All structures associated with the presented work are available from the NOMAD repository (DOI:10.17172/NOMAD/2026.02.19-1). 
 
\section{Author Contribution}

The project was conceptualised by AH and OTB. Calculations and analysis were performed by OTB.  Supervision was provided by AH. The article was drafted and critical revised by all authors, and all have provided approval of publication.

\section{Acknowledgments}

AH and OTB acknowledge funding received from European Union's Horizon 2020 research and innovation programme under grant agreements No. 953163 and No. 953167 and The authors acknowledge the use of the UCL Kathleen High Performance Computing Facility (Kathleen@UCL) and associated support services, in the completion of this work.

\section{Conflicts of Interest}

There are no conflicts to declare.

\break

\bibliography{references}

\begin{thebibliography}{71}%
\makeatletter
\providecommand \@ifxundefined [1]{%
 \@ifx{#1\undefined}
}%
\providecommand \@ifnum [1]{%
 \ifnum #1\expandafter \@firstoftwo
 \else \expandafter \@secondoftwo
 \fi
}%
\providecommand \@ifx [1]{%
 \ifx #1\expandafter \@firstoftwo
 \else \expandafter \@secondoftwo
 \fi
}%
\providecommand \natexlab [1]{#1}%
\providecommand \enquote  [1]{``#1''}%
\providecommand \bibnamefont  [1]{#1}%
\providecommand \bibfnamefont [1]{#1}%
\providecommand \citenamefont [1]{#1}%
\providecommand \href@noop [0]{\@secondoftwo}%
\providecommand \href [0]{\begingroup \@sanitize@url \@href}%
\providecommand \@href[1]{\@@startlink{#1}\@@href}%
\providecommand \@@href[1]{\endgroup#1\@@endlink}%
\providecommand \@sanitize@url [0]{\catcode `\\12\catcode `\$12\catcode `\&12\catcode `\#12\catcode `\^12\catcode `\_12\catcode `\%12\relax}%
\providecommand \@@startlink[1]{}%
\providecommand \@@endlink[0]{}%
\providecommand \url  [0]{\begingroup\@sanitize@url \@url }%
\providecommand \@url [1]{\endgroup\@href {#1}{\urlprefix }}%
\providecommand \urlprefix  [0]{URL }%
\providecommand \Eprint [0]{\href }%
\providecommand \doibase [0]{https://doi.org/}%
\providecommand \selectlanguage [0]{\@gobble}%
\providecommand \bibinfo  [0]{\@secondoftwo}%
\providecommand \bibfield  [0]{\@secondoftwo}%
\providecommand \translation [1]{[#1]}%
\providecommand \BibitemOpen [0]{}%
\providecommand \bibitemStop [0]{}%
\providecommand \bibitemNoStop [0]{.\EOS\space}%
\providecommand \EOS [0]{\spacefactor3000\relax}%
\providecommand \BibitemShut  [1]{\csname bibitem#1\endcsname}%
\let\auto@bib@innerbib\@empty
\bibitem [{\citenamefont {Yamada}\ \emph {et~al.}(1987)\citenamefont {Yamada}, \citenamefont {Ohno}, \citenamefont {Akahira}, \citenamefont {Nishiuchi}, \citenamefont {Nagata},\ and\ \citenamefont {Takao}}]{yamada_high_1987}%
  \BibitemOpen
  \bibfield  {author} {\bibinfo {author} {\bibfnamefont {N.}~\bibnamefont {Yamada}}, \bibinfo {author} {\bibfnamefont {E.}~\bibnamefont {Ohno}}, \bibinfo {author} {\bibfnamefont {N.}~\bibnamefont {Akahira}}, \bibinfo {author} {\bibfnamefont {K.}~\bibnamefont {Nishiuchi}}, \bibinfo {author} {\bibfnamefont {K.}~\bibnamefont {Nagata}},\ and\ \bibinfo {author} {\bibfnamefont {M.}~\bibnamefont {Takao}},\ }\href {https://doi.org/10.7567/JJAPS.26S4.61} {\bibfield  {journal} {\bibinfo  {journal} {Japanese Journal of Applied Physics}\ }\textbf {\bibinfo {volume} {26}},\ \bibinfo {pages} {61} (\bibinfo {year} {1987})}\BibitemShut {NoStop}%
\bibitem [{\citenamefont {Kolobov}(2024)}]{kolobov_possible_2024}%
  \BibitemOpen
  \bibfield  {author} {\bibinfo {author} {\bibfnamefont {A.~V.}\ \bibnamefont {Kolobov}},\ }\href {https://doi.org/10.1002/pssb.202300514} {\bibfield  {journal} {\bibinfo  {journal} {physica status solidi (b)}\ }\textbf {\bibinfo {volume} {261}},\ \bibinfo {pages} {2300514} (\bibinfo {year} {2024})}\BibitemShut {NoStop}%
\bibitem [{\citenamefont {Wuttig}\ and\ \citenamefont {Yamada}(2007)}]{wuttig_phase-change_2007}%
  \BibitemOpen
  \bibfield  {author} {\bibinfo {author} {\bibfnamefont {M.}~\bibnamefont {Wuttig}}\ and\ \bibinfo {author} {\bibfnamefont {N.}~\bibnamefont {Yamada}},\ }\href {https://doi.org/10.1038/nmat2009} {\bibfield  {journal} {\bibinfo  {journal} {Nature Materials}\ }\textbf {\bibinfo {volume} {6}},\ \bibinfo {pages} {824} (\bibinfo {year} {2007})}\BibitemShut {NoStop}%
\bibitem [{\citenamefont {Ovshinsky}(1968)}]{ovshinsky1968reversible}%
  \BibitemOpen
  \bibfield  {author} {\bibinfo {author} {\bibfnamefont {S.~R.}\ \bibnamefont {Ovshinsky}},\ }\href@noop {} {\bibfield  {journal} {\bibinfo  {journal} {Physical Review Letters}\ }\textbf {\bibinfo {volume} {21}},\ \bibinfo {pages} {1450} (\bibinfo {year} {1968})}\BibitemShut {NoStop}%
\bibitem [{\citenamefont {Lee}\ and\ \citenamefont {Jhi}(2008)}]{lee_ab_2008}%
  \BibitemOpen
  \bibfield  {author} {\bibinfo {author} {\bibfnamefont {G.}~\bibnamefont {Lee}}\ and\ \bibinfo {author} {\bibfnamefont {S.-H.}\ \bibnamefont {Jhi}},\ }\href {https://doi.org/10.1103/PhysRevB.77.153201} {\bibfield  {journal} {\bibinfo  {journal} {Physical Review B}\ }\textbf {\bibinfo {volume} {77}},\ \bibinfo {pages} {153201} (\bibinfo {year} {2008})}\BibitemShut {NoStop}%
\bibitem [{\citenamefont {Shportko}\ \emph {et~al.}(2008)\citenamefont {Shportko}, \citenamefont {Kremers}, \citenamefont {Woda} \emph {et~al.}}]{Shportko2008}%
  \BibitemOpen
  \bibfield  {author} {\bibinfo {author} {\bibfnamefont {K.}~\bibnamefont {Shportko}}, \bibinfo {author} {\bibfnamefont {S.}~\bibnamefont {Kremers}}, \bibinfo {author} {\bibfnamefont {M.}~\bibnamefont {Woda}}, \emph {et~al.},\ }\href {https://doi.org/10.1038/nmat2226} {\bibfield  {journal} {\bibinfo  {journal} {Nature Materials}\ }\textbf {\bibinfo {volume} {7}},\ \bibinfo {pages} {653} (\bibinfo {year} {2008})}\BibitemShut {NoStop}%
\bibitem [{\citenamefont {Xiong}\ \emph {et~al.}(2011)\citenamefont {Xiong}, \citenamefont {Liao}, \citenamefont {Estrada},\ and\ \citenamefont {Pop}}]{xiong_low-power_2011}%
  \BibitemOpen
  \bibfield  {author} {\bibinfo {author} {\bibfnamefont {F.}~\bibnamefont {Xiong}}, \bibinfo {author} {\bibfnamefont {A.~D.}\ \bibnamefont {Liao}}, \bibinfo {author} {\bibfnamefont {D.}~\bibnamefont {Estrada}},\ and\ \bibinfo {author} {\bibfnamefont {E.}~\bibnamefont {Pop}},\ }\href {https://doi.org/10.1126/science.1201938} {\bibfield  {journal} {\bibinfo  {journal} {Science}\ }\textbf {\bibinfo {volume} {332}},\ \bibinfo {pages} {568} (\bibinfo {year} {2011})}\BibitemShut {NoStop}%
\bibitem [{\citenamefont {Lankhorst}\ \emph {et~al.}(2005)\citenamefont {Lankhorst}, \citenamefont {Ketelaars},\ and\ \citenamefont {Wolters}}]{Lankhorst2005}%
  \BibitemOpen
  \bibfield  {author} {\bibinfo {author} {\bibfnamefont {M.~H.}\ \bibnamefont {Lankhorst}}, \bibinfo {author} {\bibfnamefont {B.~W.}\ \bibnamefont {Ketelaars}},\ and\ \bibinfo {author} {\bibfnamefont {R.~A.}\ \bibnamefont {Wolters}},\ }\href {https://doi.org/10.1038/nmat1350} {\bibfield  {journal} {\bibinfo  {journal} {Nature Materials}\ }\textbf {\bibinfo {volume} {4}},\ \bibinfo {pages} {347} (\bibinfo {year} {2005})}\BibitemShut {NoStop}%
\bibitem [{\citenamefont {Guo}\ \emph {et~al.}(2019)\citenamefont {Guo}, \citenamefont {Sarangan},\ and\ \citenamefont {Agha}}]{guo_review_2019}%
  \BibitemOpen
  \bibfield  {author} {\bibinfo {author} {\bibfnamefont {P.}~\bibnamefont {Guo}}, \bibinfo {author} {\bibfnamefont {A.~M.}\ \bibnamefont {Sarangan}},\ and\ \bibinfo {author} {\bibfnamefont {I.}~\bibnamefont {Agha}},\ }\href {https://doi.org/10.3390/app9030530} {\bibfield  {journal} {\bibinfo  {journal} {Applied Sciences}\ }\textbf {\bibinfo {volume} {9}},\ \bibinfo {pages} {530} (\bibinfo {year} {2019})}\BibitemShut {NoStop}%
\bibitem [{\citenamefont {D'Arrigo}\ \emph {et~al.}(2018)\citenamefont {D'Arrigo}, \citenamefont {Mio}, \citenamefont {Favaro}, \citenamefont {Calabretta}, \citenamefont {Sitta}, \citenamefont {Sciuto}, \citenamefont {Russo}, \citenamefont {Calì}, \citenamefont {Oliveri},\ and\ \citenamefont {Rimini}}]{darrigo_mechanical_2018}%
  \BibitemOpen
  \bibfield  {author} {\bibinfo {author} {\bibfnamefont {G.}~\bibnamefont {D'Arrigo}}, \bibinfo {author} {\bibfnamefont {A.}~\bibnamefont {Mio}}, \bibinfo {author} {\bibfnamefont {G.}~\bibnamefont {Favaro}}, \bibinfo {author} {\bibfnamefont {M.}~\bibnamefont {Calabretta}}, \bibinfo {author} {\bibfnamefont {A.}~\bibnamefont {Sitta}}, \bibinfo {author} {\bibfnamefont {A.}~\bibnamefont {Sciuto}}, \bibinfo {author} {\bibfnamefont {M.}~\bibnamefont {Russo}}, \bibinfo {author} {\bibfnamefont {M.}~\bibnamefont {Calì}}, \bibinfo {author} {\bibfnamefont {M.}~\bibnamefont {Oliveri}},\ and\ \bibinfo {author} {\bibfnamefont {E.}~\bibnamefont {Rimini}},\ }\href {https://doi.org/10.1016/j.surfcoat.2018.02.050} {\bibfield  {journal} {\bibinfo  {journal} {Surface and Coatings Technology}\ }\textbf {\bibinfo {volume} {355}},\ \bibinfo {pages} {227} (\bibinfo {year} {2018})}\BibitemShut {NoStop}%
\bibitem [{\citenamefont {Frumar}\ \emph {et~al.}(9 08)\citenamefont {Frumar}, \citenamefont {Kohoutek}, \citenamefont {Prikryl}, \citenamefont {Orava},\ and\ \citenamefont {Wagner}}]{frumar_atomic_2009}%
  \BibitemOpen
  \bibfield  {author} {\bibinfo {author} {\bibfnamefont {M.}~\bibnamefont {Frumar}}, \bibinfo {author} {\bibfnamefont {T.}~\bibnamefont {Kohoutek}}, \bibinfo {author} {\bibfnamefont {J.}~\bibnamefont {Prikryl}}, \bibinfo {author} {\bibfnamefont {J.}~\bibnamefont {Orava}},\ and\ \bibinfo {author} {\bibfnamefont {T.}~\bibnamefont {Wagner}},\ }\href {https://doi.org/10.1002/pssb.200982021} {\bibfield  {journal} {\bibinfo  {journal} {Physica Status Solidi B}\ }\textbf {\bibinfo {volume} {246}},\ \bibinfo {pages} {1871} (\bibinfo {year} {2009-08})}\BibitemShut {NoStop}%
\bibitem [{\citenamefont {Naito}\ \emph {et~al.}(2004)\citenamefont {Naito}, \citenamefont {Ishimaru}, \citenamefont {Hirotsu},\ and\ \citenamefont {Takashima}}]{naito_local_2004}%
  \BibitemOpen
  \bibfield  {author} {\bibinfo {author} {\bibfnamefont {M.}~\bibnamefont {Naito}}, \bibinfo {author} {\bibfnamefont {M.}~\bibnamefont {Ishimaru}}, \bibinfo {author} {\bibfnamefont {Y.}~\bibnamefont {Hirotsu}},\ and\ \bibinfo {author} {\bibfnamefont {M.}~\bibnamefont {Takashima}},\ }\href {https://pubs.aip.org/jap/article/95/12/8130/778505/Local-structure-analysis-of-Ge-Sb-Te-phase-change} {\bibfield  {journal} {\bibinfo  {journal} {Journal of Applied Physics}\ }\textbf {\bibinfo {volume} {95}},\ \bibinfo {pages} {8130} (\bibinfo {year} {2004})}\BibitemShut {NoStop}%
\bibitem [{\citenamefont {Naito}\ \emph {et~al.}(2010)\citenamefont {Naito}, \citenamefont {Ishimaru}, \citenamefont {Hirotsu}, \citenamefont {Kojima},\ and\ \citenamefont {Yamada}}]{naito_direct_2010}%
  \BibitemOpen
  \bibfield  {author} {\bibinfo {author} {\bibfnamefont {M.}~\bibnamefont {Naito}}, \bibinfo {author} {\bibfnamefont {M.}~\bibnamefont {Ishimaru}}, \bibinfo {author} {\bibfnamefont {Y.}~\bibnamefont {Hirotsu}}, \bibinfo {author} {\bibfnamefont {R.}~\bibnamefont {Kojima}},\ and\ \bibinfo {author} {\bibfnamefont {N.}~\bibnamefont {Yamada}},\ }\href {https://pubs.aip.org/jap/article/107/10/103507/146997/Direct-observations-of-Ge2Sb2Te5-recording-marks} {\bibfield  {journal} {\bibinfo  {journal} {Journal of Applied Physics}\ }\textbf {\bibinfo {volume} {107}} (\bibinfo {year} {2010})}\BibitemShut {NoStop}%
\bibitem [{\citenamefont {Bo}\ \emph {et~al.}(2004)\citenamefont {Bo}, \citenamefont {Zhi-Tang}, \citenamefont {Ting}, \citenamefont {Song-Lin},\ and\ \citenamefont {Bomy}}]{bo_raman_2004}%
  \BibitemOpen
  \bibfield  {author} {\bibinfo {author} {\bibfnamefont {L.}~\bibnamefont {Bo}}, \bibinfo {author} {\bibfnamefont {S.}~\bibnamefont {Zhi-Tang}}, \bibinfo {author} {\bibfnamefont {Z.}~\bibnamefont {Ting}}, \bibinfo {author} {\bibfnamefont {F.}~\bibnamefont {Song-Lin}},\ and\ \bibinfo {author} {\bibfnamefont {C.}~\bibnamefont {Bomy}},\ }\href {https://doi.org/10.1088/1009-1963/13/11/033} {\bibfield  {journal} {\bibinfo  {journal} {Chinese Phys.}\ }\textbf {\bibinfo {volume} {13}},\ \bibinfo {pages} {1947} (\bibinfo {year} {2004})}\BibitemShut {NoStop}%
\bibitem [{\citenamefont {Kolobov}\ \emph {et~al.}(2004)\citenamefont {Kolobov}, \citenamefont {Fons}, \citenamefont {Frenkel}, \citenamefont {Ankudinov}, \citenamefont {Tominaga},\ and\ \citenamefont {Uruga}}]{kolobov2004understanding}%
  \BibitemOpen
  \bibfield  {author} {\bibinfo {author} {\bibfnamefont {A.~V.}\ \bibnamefont {Kolobov}}, \bibinfo {author} {\bibfnamefont {P.}~\bibnamefont {Fons}}, \bibinfo {author} {\bibfnamefont {A.~I.}\ \bibnamefont {Frenkel}}, \bibinfo {author} {\bibfnamefont {A.~L.}\ \bibnamefont {Ankudinov}}, \bibinfo {author} {\bibfnamefont {J.}~\bibnamefont {Tominaga}},\ and\ \bibinfo {author} {\bibfnamefont {T.}~\bibnamefont {Uruga}},\ }\href@noop {} {\bibfield  {journal} {\bibinfo  {journal} {Nature materials}\ }\textbf {\bibinfo {volume} {3}},\ \bibinfo {pages} {703} (\bibinfo {year} {2004})}\BibitemShut {NoStop}%
\bibitem [{\citenamefont {Bala}\ \emph {et~al.}(2023)\citenamefont {Bala}, \citenamefont {Khan}, \citenamefont {Singh}, \citenamefont {Singh}, \citenamefont {Singh},\ and\ \citenamefont {Thakur}}]{bala_recent_2023}%
  \BibitemOpen
  \bibfield  {author} {\bibinfo {author} {\bibfnamefont {N.}~\bibnamefont {Bala}}, \bibinfo {author} {\bibfnamefont {B.}~\bibnamefont {Khan}}, \bibinfo {author} {\bibfnamefont {K.}~\bibnamefont {Singh}}, \bibinfo {author} {\bibfnamefont {P.}~\bibnamefont {Singh}}, \bibinfo {author} {\bibfnamefont {A.~P.}\ \bibnamefont {Singh}},\ and\ \bibinfo {author} {\bibfnamefont {A.}~\bibnamefont {Thakur}},\ }\href {https://doi.org/10.1039/D2MA01047J} {\bibfield  {journal} {\bibinfo  {journal} {Materials Advances}\ }\textbf {\bibinfo {volume} {4}},\ \bibinfo {pages} {747} (\bibinfo {year} {2023})}\BibitemShut {NoStop}%
\bibitem [{\citenamefont {Petrov}\ \emph {et~al.}(1968)\citenamefont {Petrov}, \citenamefont {Imamov},\ and\ \citenamefont {Pinsker}}]{petrov1968electron}%
  \BibitemOpen
  \bibfield  {author} {\bibinfo {author} {\bibfnamefont {I.}~\bibnamefont {Petrov}}, \bibinfo {author} {\bibfnamefont {R.}~\bibnamefont {Imamov}},\ and\ \bibinfo {author} {\bibfnamefont {Z.}~\bibnamefont {Pinsker}},\ }\href@noop {} {\bibfield  {journal} {\bibinfo  {journal} {Sov Phys Crystallogr}\ }\textbf {\bibinfo {volume} {13}},\ \bibinfo {pages} {339} (\bibinfo {year} {1968})}\BibitemShut {NoStop}%
\bibitem [{\citenamefont {Kooi}\ and\ \citenamefont {De~Hosson}(2002)}]{kooi_electron_2002}%
  \BibitemOpen
  \bibfield  {author} {\bibinfo {author} {\bibfnamefont {B.~J.}\ \bibnamefont {Kooi}}\ and\ \bibinfo {author} {\bibfnamefont {J.~T.~M.}\ \bibnamefont {De~Hosson}},\ }\href {https://doi.org/10.1063/1.1502915} {\bibfield  {journal} {\bibinfo  {journal} {Journal of Applied Physics}\ }\textbf {\bibinfo {volume} {92}},\ \bibinfo {pages} {3584} (\bibinfo {year} {2002})}\BibitemShut {NoStop}%
\bibitem [{\citenamefont {Zhou}\ \emph {et~al.}(2012)\citenamefont {Zhou}, \citenamefont {Sun}, \citenamefont {Pan}, \citenamefont {Song},\ and\ \citenamefont {Ahuja}}]{zhou_ab_2012}%
  \BibitemOpen
  \bibfield  {author} {\bibinfo {author} {\bibfnamefont {J.}~\bibnamefont {Zhou}}, \bibinfo {author} {\bibfnamefont {Z.}~\bibnamefont {Sun}}, \bibinfo {author} {\bibfnamefont {Y.}~\bibnamefont {Pan}}, \bibinfo {author} {\bibfnamefont {Z.}~\bibnamefont {Song}},\ and\ \bibinfo {author} {\bibfnamefont {R.}~\bibnamefont {Ahuja}},\ }\href {https://doi.org/10.1016/j.matchemphys.2012.01.001} {\bibfield  {journal} {\bibinfo  {journal} {Materials Chemistry and Physics}\ }\textbf {\bibinfo {volume} {133}},\ \bibinfo {pages} {159} (\bibinfo {year} {2012})}\BibitemShut {NoStop}%
\bibitem [{\citenamefont {Sun}\ \emph {et~al.}(2006)\citenamefont {Sun}, \citenamefont {Zhou},\ and\ \citenamefont {Ahuja}}]{ZhouPhysRevLett.96.055507}%
  \BibitemOpen
  \bibfield  {author} {\bibinfo {author} {\bibfnamefont {Z.}~\bibnamefont {Sun}}, \bibinfo {author} {\bibfnamefont {J.}~\bibnamefont {Zhou}},\ and\ \bibinfo {author} {\bibfnamefont {R.}~\bibnamefont {Ahuja}},\ }\href {https://doi.org/10.1103/PhysRevLett.96.055507} {\bibfield  {journal} {\bibinfo  {journal} {Phys. Rev. Lett.}\ }\textbf {\bibinfo {volume} {96}},\ \bibinfo {pages} {055507} (\bibinfo {year} {2006})}\BibitemShut {NoStop}%
\bibitem [{\citenamefont {Lyeo}\ \emph {et~al.}(2006)\citenamefont {Lyeo}, \citenamefont {Cahill}, \citenamefont {Lee}, \citenamefont {Abelson}, \citenamefont {Kwon}, \citenamefont {Kim}, \citenamefont {Bishop},\ and\ \citenamefont {Cheong}}]{lyeo_thermal_2006}%
  \BibitemOpen
  \bibfield  {author} {\bibinfo {author} {\bibfnamefont {H.-K.}\ \bibnamefont {Lyeo}}, \bibinfo {author} {\bibfnamefont {D.~G.}\ \bibnamefont {Cahill}}, \bibinfo {author} {\bibfnamefont {B.-S.}\ \bibnamefont {Lee}}, \bibinfo {author} {\bibfnamefont {J.~R.}\ \bibnamefont {Abelson}}, \bibinfo {author} {\bibfnamefont {M.-H.}\ \bibnamefont {Kwon}}, \bibinfo {author} {\bibfnamefont {K.-B.}\ \bibnamefont {Kim}}, \bibinfo {author} {\bibfnamefont {S.~G.}\ \bibnamefont {Bishop}},\ and\ \bibinfo {author} {\bibfnamefont {B.-k.}\ \bibnamefont {Cheong}},\ }\href {https://doi.org/10.1063/1.2359354} {\bibfield  {journal} {\bibinfo  {journal} {Applied Physics Letters}\ }\textbf {\bibinfo {volume} {89}},\ \bibinfo {pages} {151904} (\bibinfo {year} {2006})}\BibitemShut {NoStop}%
\bibitem [{\citenamefont {Duong}\ \emph {et~al.}(2021)\citenamefont {Duong}, \citenamefont {Bouzid}, \citenamefont {Massobrio}, \citenamefont {Ori}, \citenamefont {Boero},\ and\ \citenamefont {Martin}}]{duong_first-principles_2021}%
  \BibitemOpen
  \bibfield  {author} {\bibinfo {author} {\bibfnamefont {T.~Q.}\ \bibnamefont {Duong}}, \bibinfo {author} {\bibfnamefont {A.}~\bibnamefont {Bouzid}}, \bibinfo {author} {\bibfnamefont {C.}~\bibnamefont {Massobrio}}, \bibinfo {author} {\bibfnamefont {G.}~\bibnamefont {Ori}}, \bibinfo {author} {\bibfnamefont {M.}~\bibnamefont {Boero}},\ and\ \bibinfo {author} {\bibfnamefont {E.}~\bibnamefont {Martin}},\ }\href {https://doi.org/10.1039/D0RA10408F} {\bibfield  {journal} {\bibinfo  {journal} {{RSC} Advances}\ }\textbf {\bibinfo {volume} {11}},\ \bibinfo {pages} {10747} (\bibinfo {year} {2021})}\BibitemShut {NoStop}%
\bibitem [{\citenamefont {Ghosh}\ \emph {et~al.}(2023)\citenamefont {Ghosh}, \citenamefont {Kusiak},\ and\ \citenamefont {Battaglia}}]{ghosh_significant_2023}%
  \BibitemOpen
  \bibfield  {author} {\bibinfo {author} {\bibfnamefont {K.}~\bibnamefont {Ghosh}}, \bibinfo {author} {\bibfnamefont {A.}~\bibnamefont {Kusiak}},\ and\ \bibinfo {author} {\bibfnamefont {J.-L.}\ \bibnamefont {Battaglia}},\ }\href {https://doi.org/10.1103/PhysRevB.108.214309} {\bibfield  {journal} {\bibinfo  {journal} {Physical Review B}\ }\textbf {\bibinfo {volume} {108}},\ \bibinfo {pages} {214309} (\bibinfo {year} {2023})}\BibitemShut {NoStop}%
\bibitem [{\citenamefont {Deringer}\ and\ \citenamefont {Dronskowski}(2013)}]{deringer_2013_dft}%
  \BibitemOpen
  \bibfield  {author} {\bibinfo {author} {\bibfnamefont {V.~L.}\ \bibnamefont {Deringer}}\ and\ \bibinfo {author} {\bibfnamefont {R.}~\bibnamefont {Dronskowski}},\ }\href {https://doi.org/10.1021/jp401400k} {\bibfield  {journal} {\bibinfo  {journal} {The Journal of Physical Chemistry C}\ }\textbf {\bibinfo {volume} {117}},\ \bibinfo {pages} {15075} (\bibinfo {year} {2013})}\BibitemShut {NoStop}%
\bibitem [{\citenamefont {Ahn}\ \emph {et~al.}(2020)\citenamefont {Ahn}, \citenamefont {sik Jeong}, \citenamefont {Park}, \citenamefont {Jung}, \citenamefont {Han}, \citenamefont {Yang}, \citenamefont {Kim}, \citenamefont {Park},\ and\ \citenamefont {Cho}}]{AHN2020807}%
  \BibitemOpen
  \bibfield  {author} {\bibinfo {author} {\bibfnamefont {M.}~\bibnamefont {Ahn}}, \bibinfo {author} {\bibfnamefont {K.}~\bibnamefont {sik Jeong}}, \bibinfo {author} {\bibfnamefont {S.}~\bibnamefont {Park}}, \bibinfo {author} {\bibfnamefont {H.}~\bibnamefont {Jung}}, \bibinfo {author} {\bibfnamefont {J.}~\bibnamefont {Han}}, \bibinfo {author} {\bibfnamefont {W.}~\bibnamefont {Yang}}, \bibinfo {author} {\bibfnamefont {D.}~\bibnamefont {Kim}}, \bibinfo {author} {\bibfnamefont {J.}~\bibnamefont {Park}},\ and\ \bibinfo {author} {\bibfnamefont {M.-H.}\ \bibnamefont {Cho}},\ }\href {https://doi.org/https://doi.org/10.1016/j.cap.2020.03.019} {\bibfield  {journal} {\bibinfo  {journal} {Current Applied Physics}\ }\textbf {\bibinfo {volume} {20}},\ \bibinfo {pages} {807} (\bibinfo {year} {2020})}\BibitemShut {NoStop}%
\bibitem [{\citenamefont {Beynon}\ \emph {et~al.}(2023)\citenamefont {Beynon}, \citenamefont {Owens}, \citenamefont {Carbogno},\ and\ \citenamefont {Logsdail}}]{doi:10.1021/acs.jpcc.3c02863}%
  \BibitemOpen
  \bibfield  {author} {\bibinfo {author} {\bibfnamefont {O.~T.}\ \bibnamefont {Beynon}}, \bibinfo {author} {\bibfnamefont {A.}~\bibnamefont {Owens}}, \bibinfo {author} {\bibfnamefont {C.}~\bibnamefont {Carbogno}},\ and\ \bibinfo {author} {\bibfnamefont {A.~J.}\ \bibnamefont {Logsdail}},\ }\href@noop {} {\bibfield  {journal} {\bibinfo  {journal} {The Journal of Physical Chemistry C}\ }\textbf {\bibinfo {volume} {127}},\ \bibinfo {pages} {16030} (\bibinfo {year} {2023})}\BibitemShut {NoStop}%
\bibitem [{\citenamefont {Scott}\ \emph {et~al.}(2020)\citenamefont {Scott}, \citenamefont {Ziade}, \citenamefont {Saltonstall}, \citenamefont {McDonald}, \citenamefont {Rodriguez}, \citenamefont {Hopkins}, \citenamefont {Beechem},\ and\ \citenamefont {Adams}}]{10.1063/5.0023476}%
  \BibitemOpen
  \bibfield  {author} {\bibinfo {author} {\bibfnamefont {E.~A.}\ \bibnamefont {Scott}}, \bibinfo {author} {\bibfnamefont {E.}~\bibnamefont {Ziade}}, \bibinfo {author} {\bibfnamefont {C.~B.}\ \bibnamefont {Saltonstall}}, \bibinfo {author} {\bibfnamefont {A.~E.}\ \bibnamefont {McDonald}}, \bibinfo {author} {\bibfnamefont {M.~A.}\ \bibnamefont {Rodriguez}}, \bibinfo {author} {\bibfnamefont {P.~E.}\ \bibnamefont {Hopkins}}, \bibinfo {author} {\bibfnamefont {T.~E.}\ \bibnamefont {Beechem}},\ and\ \bibinfo {author} {\bibfnamefont {D.~P.}\ \bibnamefont {Adams}},\ }\href {https://doi.org/10.1063/5.0023476} {\bibfield  {journal} {\bibinfo  {journal} {Journal of Applied Physics}\ }\textbf {\bibinfo {volume} {128}},\ \bibinfo {pages} {155106} (\bibinfo {year} {2020})}\BibitemShut {NoStop}%
\bibitem [{\citenamefont {Knoop}\ \emph {et~al.}(2020{\natexlab{a}})\citenamefont {Knoop}, \citenamefont {Purcell}, \citenamefont {Scheffler},\ and\ \citenamefont {Carbogno}}]{PhysRevMaterials.4.083809}%
  \BibitemOpen
  \bibfield  {author} {\bibinfo {author} {\bibfnamefont {F.}~\bibnamefont {Knoop}}, \bibinfo {author} {\bibfnamefont {T.~A.~R.}\ \bibnamefont {Purcell}}, \bibinfo {author} {\bibfnamefont {M.}~\bibnamefont {Scheffler}},\ and\ \bibinfo {author} {\bibfnamefont {C.}~\bibnamefont {Carbogno}},\ }\href {https://doi.org/10.1103/PhysRevMaterials.4.083809} {\bibfield  {journal} {\bibinfo  {journal} {Phys. Rev. Mater.}\ }\textbf {\bibinfo {volume} {4}},\ \bibinfo {pages} {083809} (\bibinfo {year} {2020}{\natexlab{a}})}\BibitemShut {NoStop}%
\bibitem [{\citenamefont {Blum}\ \emph {et~al.}(2009)\citenamefont {Blum}, \citenamefont {Gehrke}, \citenamefont {Hanke}, \citenamefont {Havu}, \citenamefont {Havu}, \citenamefont {Ren}, \citenamefont {Reuter},\ and\ \citenamefont {Scheffler}}]{blum_ab_2009}%
  \BibitemOpen
  \bibfield  {author} {\bibinfo {author} {\bibfnamefont {V.}~\bibnamefont {Blum}}, \bibinfo {author} {\bibfnamefont {R.}~\bibnamefont {Gehrke}}, \bibinfo {author} {\bibfnamefont {F.}~\bibnamefont {Hanke}}, \bibinfo {author} {\bibfnamefont {P.}~\bibnamefont {Havu}}, \bibinfo {author} {\bibfnamefont {V.}~\bibnamefont {Havu}}, \bibinfo {author} {\bibfnamefont {X.}~\bibnamefont {Ren}}, \bibinfo {author} {\bibfnamefont {K.}~\bibnamefont {Reuter}},\ and\ \bibinfo {author} {\bibfnamefont {M.}~\bibnamefont {Scheffler}},\ }\href {https://doi.org/10.1016/j.cpc.2009.06.022} {\bibfield  {journal} {\bibinfo  {journal} {Computer Physics Communications}\ }\textbf {\bibinfo {volume} {180}},\ \bibinfo {pages} {2175} (\bibinfo {year} {2009})}\BibitemShut {NoStop}%
\bibitem [{\citenamefont {Perdew}\ \emph {et~al.}(1996)\citenamefont {Perdew}, \citenamefont {Burke},\ and\ \citenamefont {Ernzerhof}}]{perdew_generalized_1996}%
  \BibitemOpen
  \bibfield  {author} {\bibinfo {author} {\bibfnamefont {J.~P.}\ \bibnamefont {Perdew}}, \bibinfo {author} {\bibfnamefont {K.}~\bibnamefont {Burke}},\ and\ \bibinfo {author} {\bibfnamefont {M.}~\bibnamefont {Ernzerhof}},\ }\href {https://doi.org/10.1103/PhysRevLett.77.3865} {\bibfield  {journal} {\bibinfo  {journal} {Physical Review Letters}\ }\textbf {\bibinfo {volume} {77}},\ \bibinfo {pages} {3865} (\bibinfo {year} {1996})}\BibitemShut {NoStop}%
\bibitem [{\citenamefont {Perdew}\ \emph {et~al.}(2008)\citenamefont {Perdew}, \citenamefont {Ruzsinszky}, \citenamefont {Csonka}, \citenamefont {Vydrov}, \citenamefont {Scuseria}, \citenamefont {Constantin}, \citenamefont {Zhou},\ and\ \citenamefont {Burke}}]{perdew_restoring_2008}%
  \BibitemOpen
  \bibfield  {author} {\bibinfo {author} {\bibfnamefont {J.~P.}\ \bibnamefont {Perdew}}, \bibinfo {author} {\bibfnamefont {A.}~\bibnamefont {Ruzsinszky}}, \bibinfo {author} {\bibfnamefont {G.~I.}\ \bibnamefont {Csonka}}, \bibinfo {author} {\bibfnamefont {O.~A.}\ \bibnamefont {Vydrov}}, \bibinfo {author} {\bibfnamefont {G.~E.}\ \bibnamefont {Scuseria}}, \bibinfo {author} {\bibfnamefont {L.~A.}\ \bibnamefont {Constantin}}, \bibinfo {author} {\bibfnamefont {X.}~\bibnamefont {Zhou}},\ and\ \bibinfo {author} {\bibfnamefont {K.}~\bibnamefont {Burke}},\ }\href {https://doi.org/10.1103/PhysRevLett.100.136406} {\bibfield  {journal} {\bibinfo  {journal} {Physical Review Letters}\ }\textbf {\bibinfo {volume} {100}},\ \bibinfo {pages} {136406} (\bibinfo {year} {2008})}\BibitemShut {NoStop}%
\bibitem [{\citenamefont {Tkatchenko}\ and\ \citenamefont {Scheffler}(2009)}]{tkatchenko_accurate_2009}%
  \BibitemOpen
  \bibfield  {author} {\bibinfo {author} {\bibfnamefont {A.}~\bibnamefont {Tkatchenko}}\ and\ \bibinfo {author} {\bibfnamefont {M.}~\bibnamefont {Scheffler}},\ }\href {https://doi.org/10.1103/PhysRevLett.102.073005} {\bibfield  {journal} {\bibinfo  {journal} {Physical Review Letters}\ }\textbf {\bibinfo {volume} {102}},\ \bibinfo {pages} {073005} (\bibinfo {year} {2009})}\BibitemShut {NoStop}%
\bibitem [{\citenamefont {Monkhorst}\ and\ \citenamefont {Pack}(1976)}]{monkhorst_special_1976}%
  \BibitemOpen
  \bibfield  {author} {\bibinfo {author} {\bibfnamefont {H.~J.}\ \bibnamefont {Monkhorst}}\ and\ \bibinfo {author} {\bibfnamefont {J.~D.}\ \bibnamefont {Pack}},\ }\href {https://doi.org/10.1103/PhysRevB.13.5188} {\bibfield  {journal} {\bibinfo  {journal} {Physical Review B}\ }\textbf {\bibinfo {volume} {13}},\ \bibinfo {pages} {5188} (\bibinfo {year} {1976})}\BibitemShut {NoStop}%
\bibitem [{\citenamefont {van Lenthe}\ \emph {et~al.}(4 12)\citenamefont {van Lenthe}, \citenamefont {Baerends},\ and\ \citenamefont {Snijders}}]{van_lenthe_relativistic_1994}%
  \BibitemOpen
  \bibfield  {author} {\bibinfo {author} {\bibfnamefont {E.}~\bibnamefont {van Lenthe}}, \bibinfo {author} {\bibfnamefont {E.~J.}\ \bibnamefont {Baerends}},\ and\ \bibinfo {author} {\bibfnamefont {J.~G.}\ \bibnamefont {Snijders}},\ }\href {https://doi.org/10.1063/1.467943} {\bibfield  {journal} {\bibinfo  {journal} {The Journal of Chemical Physics}\ }\textbf {\bibinfo {volume} {101}},\ \bibinfo {pages} {9783} (\bibinfo {year} {1994-12})}\BibitemShut {NoStop}%
\bibitem [{\citenamefont {Hjorth~Larsen}\ \emph {et~al.}(2017)\citenamefont {Hjorth~Larsen}, \citenamefont {Jørgen~Mortensen}, \citenamefont {Blomqvist}, \citenamefont {Castelli}, \citenamefont {Christensen}, \citenamefont {Dułak}, \citenamefont {Friis}, \citenamefont {Groves}, \citenamefont {Hammer}, \citenamefont {Hargus}, \citenamefont {Hermes}, \citenamefont {Jennings}, \citenamefont {Bjerre~Jensen}, \citenamefont {Kermode}, \citenamefont {Kitchin}, \citenamefont {Leonhard~Kolsbjerg}, \citenamefont {Kubal}, \citenamefont {Kaasbjerg}, \citenamefont {Lysgaard}, \citenamefont {Bergmann~Maronsson}, \citenamefont {Maxson}, \citenamefont {Olsen}, \citenamefont {Pastewka}, \citenamefont {Peterson}, \citenamefont {Rostgaard}, \citenamefont {Schiøtz}, \citenamefont {Schütt}, \citenamefont {Strange}, \citenamefont {Thygesen}, \citenamefont {Vegge}, \citenamefont {Vilhelmsen}, \citenamefont {Walter}, \citenamefont {Zeng},\ and\ \citenamefont {Jacobsen}}]{hjorth_larsen_atomic_2017}%
  \BibitemOpen
  \bibfield  {author} {\bibinfo {author} {\bibfnamefont {A.}~\bibnamefont {Hjorth~Larsen}}, \bibinfo {author} {\bibfnamefont {J.}~\bibnamefont {Jørgen~Mortensen}}, \bibinfo {author} {\bibfnamefont {J.}~\bibnamefont {Blomqvist}}, \bibinfo {author} {\bibfnamefont {I.~E.}\ \bibnamefont {Castelli}}, \bibinfo {author} {\bibfnamefont {R.}~\bibnamefont {Christensen}}, \bibinfo {author} {\bibfnamefont {M.}~\bibnamefont {Dułak}}, \bibinfo {author} {\bibfnamefont {J.}~\bibnamefont {Friis}}, \bibinfo {author} {\bibfnamefont {M.~N.}\ \bibnamefont {Groves}}, \bibinfo {author} {\bibfnamefont {B.}~\bibnamefont {Hammer}}, \bibinfo {author} {\bibfnamefont {C.}~\bibnamefont {Hargus}}, \bibinfo {author} {\bibfnamefont {E.~D.}\ \bibnamefont {Hermes}}, \bibinfo {author} {\bibfnamefont {P.~C.}\ \bibnamefont {Jennings}}, \bibinfo {author} {\bibfnamefont {P.}~\bibnamefont {Bjerre~Jensen}}, \bibinfo {author} {\bibfnamefont {J.}~\bibnamefont {Kermode}}, \bibinfo {author} {\bibfnamefont {J.~R.}\ \bibnamefont {Kitchin}}, \bibinfo
  {author} {\bibfnamefont {E.}~\bibnamefont {Leonhard~Kolsbjerg}}, \bibinfo {author} {\bibfnamefont {J.}~\bibnamefont {Kubal}}, \bibinfo {author} {\bibfnamefont {K.}~\bibnamefont {Kaasbjerg}}, \bibinfo {author} {\bibfnamefont {S.}~\bibnamefont {Lysgaard}}, \bibinfo {author} {\bibfnamefont {J.}~\bibnamefont {Bergmann~Maronsson}}, \bibinfo {author} {\bibfnamefont {T.}~\bibnamefont {Maxson}}, \bibinfo {author} {\bibfnamefont {T.}~\bibnamefont {Olsen}}, \bibinfo {author} {\bibfnamefont {L.}~\bibnamefont {Pastewka}}, \bibinfo {author} {\bibfnamefont {A.}~\bibnamefont {Peterson}}, \bibinfo {author} {\bibfnamefont {C.}~\bibnamefont {Rostgaard}}, \bibinfo {author} {\bibfnamefont {J.}~\bibnamefont {Schiøtz}}, \bibinfo {author} {\bibfnamefont {O.}~\bibnamefont {Schütt}}, \bibinfo {author} {\bibfnamefont {M.}~\bibnamefont {Strange}}, \bibinfo {author} {\bibfnamefont {K.~S.}\ \bibnamefont {Thygesen}}, \bibinfo {author} {\bibfnamefont {T.}~\bibnamefont {Vegge}}, \bibinfo {author} {\bibfnamefont {L.}~\bibnamefont
  {Vilhelmsen}}, \bibinfo {author} {\bibfnamefont {M.}~\bibnamefont {Walter}}, \bibinfo {author} {\bibfnamefont {Z.}~\bibnamefont {Zeng}},\ and\ \bibinfo {author} {\bibfnamefont {K.~W.}\ \bibnamefont {Jacobsen}},\ }\href {https://doi.org/10.1088/1361-648X/aa680e} {\bibfield  {journal} {\bibinfo  {journal} {Journal of Physics: Condensed Matter}\ }\textbf {\bibinfo {volume} {29}},\ \bibinfo {pages} {273002} (\bibinfo {year} {2017})}\BibitemShut {NoStop}%
\bibitem [{\citenamefont {Broyden}(1970)}]{broyden_convergence_1970}%
  \BibitemOpen
  \bibfield  {author} {\bibinfo {author} {\bibfnamefont {C.~G.}\ \bibnamefont {Broyden}},\ }\href {https://doi.org/10.1093/imamat/6.1.76} {\bibfield  {journal} {\bibinfo  {journal} {{IMA} Journal of Applied Mathematics}\ }\textbf {\bibinfo {volume} {6}},\ \bibinfo {pages} {76} (\bibinfo {year} {1970})}\BibitemShut {NoStop}%
\bibitem [{\citenamefont {Fletcher}(1970)}]{fletcher_new_1970}%
  \BibitemOpen
  \bibfield  {author} {\bibinfo {author} {\bibfnamefont {R.}~\bibnamefont {Fletcher}},\ }\href {https://doi.org/10.1093/comjnl/13.3.317} {\bibfield  {journal} {\bibinfo  {journal} {The Computer Journal}\ }\textbf {\bibinfo {volume} {13}},\ \bibinfo {pages} {317} (\bibinfo {year} {1970})}\BibitemShut {NoStop}%
\bibitem [{\citenamefont {Goldfarb}(1970)}]{goldfarb_family_1970}%
  \BibitemOpen
  \bibfield  {author} {\bibinfo {author} {\bibfnamefont {D.}~\bibnamefont {Goldfarb}},\ }\href {https://doi.org/10.1090/S0025-5718-1970-0258249-6} {\bibfield  {journal} {\bibinfo  {journal} {Mathematics of Computation}\ }\textbf {\bibinfo {volume} {24}},\ \bibinfo {pages} {23} (\bibinfo {year} {1970})}\BibitemShut {NoStop}%
\bibitem [{\citenamefont {Shanno}(1970)}]{shanno_conditioning_1970}%
  \BibitemOpen
  \bibfield  {author} {\bibinfo {author} {\bibfnamefont {D.~F.}\ \bibnamefont {Shanno}},\ }\href {https://doi.org/10.1090/S0025-5718-1970-0274029-X} {\bibfield  {journal} {\bibinfo  {journal} {Mathematics of Computation}\ }\textbf {\bibinfo {volume} {24}},\ \bibinfo {pages} {647} (\bibinfo {year} {1970})}\BibitemShut {NoStop}%
\bibitem [{\citenamefont {Knoop}\ \emph {et~al.}(2020{\natexlab{b}})\citenamefont {Knoop}, \citenamefont {Purcell}, \citenamefont {Scheffler},\ and\ \citenamefont {Carbogno}}]{knoop_fhi-vibes_2020}%
  \BibitemOpen
  \bibfield  {author} {\bibinfo {author} {\bibfnamefont {F.}~\bibnamefont {Knoop}}, \bibinfo {author} {\bibfnamefont {T.}~\bibnamefont {Purcell}}, \bibinfo {author} {\bibfnamefont {M.}~\bibnamefont {Scheffler}},\ and\ \bibinfo {author} {\bibfnamefont {C.}~\bibnamefont {Carbogno}},\ }\href {https://doi.org/10.21105/joss.02671} {\bibfield  {journal} {\bibinfo  {journal} {Journal of Open Source Software}\ }\textbf {\bibinfo {volume} {5}},\ \bibinfo {pages} {2671} (\bibinfo {year} {2020}{\natexlab{b}})}\BibitemShut {NoStop}%
\bibitem [{\citenamefont {Togo}\ and\ \citenamefont {Tanaka}(2015)}]{togo_first_2015}%
  \BibitemOpen
  \bibfield  {author} {\bibinfo {author} {\bibfnamefont {A.}~\bibnamefont {Togo}}\ and\ \bibinfo {author} {\bibfnamefont {I.}~\bibnamefont {Tanaka}},\ }\href {https://doi.org/10.1016/j.scriptamat.2015.07.021} {\bibfield  {journal} {\bibinfo  {journal} {Scripta Materialia}\ }\textbf {\bibinfo {volume} {108}},\ \bibinfo {pages} {1} (\bibinfo {year} {2015})}\BibitemShut {NoStop}%
\bibitem [{\citenamefont {Eriksson}\ \emph {et~al.}(9 05)\citenamefont {Eriksson}, \citenamefont {Fransson},\ and\ \citenamefont {Erhart}}]{eriksson_hiphive_2019}%
  \BibitemOpen
  \bibfield  {author} {\bibinfo {author} {\bibfnamefont {F.}~\bibnamefont {Eriksson}}, \bibinfo {author} {\bibfnamefont {E.}~\bibnamefont {Fransson}},\ and\ \bibinfo {author} {\bibfnamefont {P.}~\bibnamefont {Erhart}},\ }\href {https://doi.org/10.1002/adts.201800184} {\bibfield  {journal} {\bibinfo  {journal} {Advanced Theory and Simulations}\ }\textbf {\bibinfo {volume} {2}},\ \bibinfo {pages} {1800184} (\bibinfo {year} {2019-05})}\BibitemShut {NoStop}%
\bibitem [{\citenamefont {Singh}\ \emph {et~al.}(2018)\citenamefont {Singh}, \citenamefont {Singh}, \citenamefont {Kaura},\ and\ \citenamefont {Tripathi}}]{SINGH2018124}%
  \BibitemOpen
  \bibfield  {author} {\bibinfo {author} {\bibfnamefont {J.}~\bibnamefont {Singh}}, \bibinfo {author} {\bibfnamefont {G.}~\bibnamefont {Singh}}, \bibinfo {author} {\bibfnamefont {A.}~\bibnamefont {Kaura}},\ and\ \bibinfo {author} {\bibfnamefont {S.}~\bibnamefont {Tripathi}},\ }\href {https://doi.org/https://doi.org/10.1016/j.jssc.2018.01.021} {\bibfield  {journal} {\bibinfo  {journal} {Journal of Solid State Chemistry}\ }\textbf {\bibinfo {volume} {260}},\ \bibinfo {pages} {124} (\bibinfo {year} {2018})}\BibitemShut {NoStop}%
\bibitem [{\citenamefont {He}\ \emph {et~al.}(2017)\citenamefont {He}, \citenamefont {Zhu}, \citenamefont {Zhou},\ and\ \citenamefont {Sun}}]{doi:10.1021/acs.inorgchem.7b01970}%
  \BibitemOpen
  \bibfield  {author} {\bibinfo {author} {\bibfnamefont {S.}~\bibnamefont {He}}, \bibinfo {author} {\bibfnamefont {L.}~\bibnamefont {Zhu}}, \bibinfo {author} {\bibfnamefont {J.}~\bibnamefont {Zhou}},\ and\ \bibinfo {author} {\bibfnamefont {Z.}~\bibnamefont {Sun}},\ }\href {https://doi.org/10.1021/acs.inorgchem.7b01970} {\bibfield  {journal} {\bibinfo  {journal} {Inorganic Chemistry}\ }\textbf {\bibinfo {volume} {56}},\ \bibinfo {pages} {11990} (\bibinfo {year} {2017})}\BibitemShut {NoStop}%
\bibitem [{\citenamefont {Sa}\ \emph {et~al.}(2014)\citenamefont {Sa}, \citenamefont {Zhou}, \citenamefont {Ahuja},\ and\ \citenamefont {Sun}}]{sa_first-principles_2014}%
  \BibitemOpen
  \bibfield  {author} {\bibinfo {author} {\bibfnamefont {B.}~\bibnamefont {Sa}}, \bibinfo {author} {\bibfnamefont {J.}~\bibnamefont {Zhou}}, \bibinfo {author} {\bibfnamefont {R.}~\bibnamefont {Ahuja}},\ and\ \bibinfo {author} {\bibfnamefont {Z.}~\bibnamefont {Sun}},\ }\href {https://doi.org/10.1016/j.commatsci.2013.09.026} {\bibfield  {journal} {\bibinfo  {journal} {Computational Materials Science}\ }\textbf {\bibinfo {volume} {82}},\ \bibinfo {pages} {66} (\bibinfo {year} {2014})}\BibitemShut {NoStop}%
\bibitem [{\citenamefont {Cooley}\ \emph {et~al.}(2020)\citenamefont {Cooley}, \citenamefont {Keilbart}, \citenamefont {Champlain}, \citenamefont {Ruppalt}, \citenamefont {Walter}, \citenamefont {Dabo},\ and\ \citenamefont {Mohney}}]{cooley_first-principles_2020}%
  \BibitemOpen
  \bibfield  {author} {\bibinfo {author} {\bibfnamefont {K.~A.}\ \bibnamefont {Cooley}}, \bibinfo {author} {\bibfnamefont {N.}~\bibnamefont {Keilbart}}, \bibinfo {author} {\bibfnamefont {J.~G.}\ \bibnamefont {Champlain}}, \bibinfo {author} {\bibfnamefont {L.~B.}\ \bibnamefont {Ruppalt}}, \bibinfo {author} {\bibfnamefont {T.~N.}\ \bibnamefont {Walter}}, \bibinfo {author} {\bibfnamefont {I.}~\bibnamefont {Dabo}},\ and\ \bibinfo {author} {\bibfnamefont {S.~E.}\ \bibnamefont {Mohney}},\ }\href {https://doi.org/10.1063/5.0029205} {\bibfield  {journal} {\bibinfo  {journal} {Journal of Applied Physics}\ }\textbf {\bibinfo {volume} {128}},\ \bibinfo {pages} {225306} (\bibinfo {year} {2020})}\BibitemShut {NoStop}%
\bibitem [{\citenamefont {Yang}\ \emph {et~al.}(2018)\citenamefont {Yang}, \citenamefont {Chen}, \citenamefont {Wang}, \citenamefont {Yan}, \citenamefont {Wan}, \citenamefont {Ke},\ and\ \citenamefont {Dai}}]{AIPAdv2018}%
  \BibitemOpen
  \bibfield  {author} {\bibinfo {author} {\bibfnamefont {F.}~\bibnamefont {Yang}}, \bibinfo {author} {\bibfnamefont {T.}~\bibnamefont {Chen}}, \bibinfo {author} {\bibfnamefont {M.}~\bibnamefont {Wang}}, \bibinfo {author} {\bibfnamefont {B.}~\bibnamefont {Yan}}, \bibinfo {author} {\bibfnamefont {L.}~\bibnamefont {Wan}}, \bibinfo {author} {\bibfnamefont {D.}~\bibnamefont {Ke}},\ and\ \bibinfo {author} {\bibfnamefont {Y.}~\bibnamefont {Dai}},\ }\href {https://doi.org/10.1063/1.5006247} {\bibfield  {journal} {\bibinfo  {journal} {AIP Advances}\ }\textbf {\bibinfo {volume} {8}},\ \bibinfo {pages} {065223} (\bibinfo {year} {2018})}\BibitemShut {NoStop}%
\bibitem [{\citenamefont {Ibarra-Hern\'andez}\ and\ \citenamefont {Raty}(2018)}]{IbarraHernPhysRevB.97.245205}%
  \BibitemOpen
  \bibfield  {author} {\bibinfo {author} {\bibfnamefont {W.}~\bibnamefont {Ibarra-Hern\'andez}}\ and\ \bibinfo {author} {\bibfnamefont {J.-Y.}\ \bibnamefont {Raty}},\ }\href {https://doi.org/10.1103/PhysRevB.97.245205} {\bibfield  {journal} {\bibinfo  {journal} {Phys. Rev. B}\ }\textbf {\bibinfo {volume} {97}},\ \bibinfo {pages} {245205} (\bibinfo {year} {2018})}\BibitemShut {NoStop}%
\bibitem [{\citenamefont {Hashibon}\ and\ \citenamefont {Els\"asser}(2011)}]{PhysRevB.84.144117}%
  \BibitemOpen
  \bibfield  {author} {\bibinfo {author} {\bibfnamefont {A.}~\bibnamefont {Hashibon}}\ and\ \bibinfo {author} {\bibfnamefont {C.}~\bibnamefont {Els\"asser}},\ }\href {https://doi.org/10.1103/PhysRevB.84.144117} {\bibfield  {journal} {\bibinfo  {journal} {Phys. Rev. B}\ }\textbf {\bibinfo {volume} {84}},\ \bibinfo {pages} {144117} (\bibinfo {year} {2011})}\BibitemShut {NoStop}%
\bibitem [{\citenamefont {Cheng}(2024)}]{PhysRevA.110.042805}%
  \BibitemOpen
  \bibfield  {author} {\bibinfo {author} {\bibfnamefont {Y.}~\bibnamefont {Cheng}},\ }\href {https://doi.org/10.1103/PhysRevA.110.042805} {\bibfield  {journal} {\bibinfo  {journal} {Phys. Rev. A}\ }\textbf {\bibinfo {volume} {110}},\ \bibinfo {pages} {042805} (\bibinfo {year} {2024})}\BibitemShut {NoStop}%
\bibitem [{\citenamefont {Karjalainen}\ \emph {et~al.}(2016)\citenamefont {Karjalainen}, \citenamefont {Sanchez-Perez}, \citenamefont {Rautiainen}, \citenamefont {Oilunkaniemi},\ and\ \citenamefont {Laitinen}}]{C6CE00451B}%
  \BibitemOpen
  \bibfield  {author} {\bibinfo {author} {\bibfnamefont {M.~M.}\ \bibnamefont {Karjalainen}}, \bibinfo {author} {\bibfnamefont {C.}~\bibnamefont {Sanchez-Perez}}, \bibinfo {author} {\bibfnamefont {J.~M.}\ \bibnamefont {Rautiainen}}, \bibinfo {author} {\bibfnamefont {R.}~\bibnamefont {Oilunkaniemi}},\ and\ \bibinfo {author} {\bibfnamefont {R.~S.}\ \bibnamefont {Laitinen}},\ }\href {https://doi.org/10.1039/C6CE00451B} {\bibfield  {journal} {\bibinfo  {journal} {CrystEngComm}\ }\textbf {\bibinfo {volume} {18}},\ \bibinfo {pages} {4538} (\bibinfo {year} {2016})}\BibitemShut {NoStop}%
\bibitem [{\citenamefont {Sa}\ \emph {et~al.}(2013)\citenamefont {Sa}, \citenamefont {Sun}, \citenamefont {Kaewmaraya}, \citenamefont {Zhou},\ and\ \citenamefont {Ahuja}}]{sa_structural_2013}%
  \BibitemOpen
  \bibfield  {author} {\bibinfo {author} {\bibfnamefont {B.}~\bibnamefont {Sa}}, \bibinfo {author} {\bibfnamefont {Z.}~\bibnamefont {Sun}}, \bibinfo {author} {\bibfnamefont {T.}~\bibnamefont {Kaewmaraya}}, \bibinfo {author} {\bibfnamefont {J.}~\bibnamefont {Zhou}},\ and\ \bibinfo {author} {\bibfnamefont {R.}~\bibnamefont {Ahuja}},\ }\href {https://doi.org/10.1166/sam.2013.1610} {\bibfield  {journal} {\bibinfo  {journal} {Science of Advanced Materials}\ }\textbf {\bibinfo {volume} {5}},\ \bibinfo {pages} {1493} (\bibinfo {year} {2013})}\BibitemShut {NoStop}%
\bibitem [{\citenamefont {Matsunaga}\ \emph {et~al.}(2004)\citenamefont {Matsunaga}, \citenamefont {Yamada},\ and\ \citenamefont {Kubota}}]{matsunaga_structures_2004}%
  \BibitemOpen
  \bibfield  {author} {\bibinfo {author} {\bibfnamefont {T.}~\bibnamefont {Matsunaga}}, \bibinfo {author} {\bibfnamefont {N.}~\bibnamefont {Yamada}},\ and\ \bibinfo {author} {\bibfnamefont {Y.}~\bibnamefont {Kubota}},\ }\href {https://doi.org/10.1107/S0108768104022906} {\bibfield  {journal} {\bibinfo  {journal} {Acta Crystallographica Section B Structural Science}\ }\textbf {\bibinfo {volume} {60}},\ \bibinfo {pages} {685} (\bibinfo {year} {2004})}\BibitemShut {NoStop}%
\bibitem [{\citenamefont {Micoulaut}(2013)}]{Micoulaut2013}%
  \BibitemOpen
  \bibfield  {author} {\bibinfo {author} {\bibfnamefont {M.}~\bibnamefont {Micoulaut}},\ }\href {https://doi.org/10.1063/1.4791715} {\bibfield  {journal} {\bibinfo  {journal} {The Journal of Chemical Physics}\ }\textbf {\bibinfo {volume} {138}},\ \bibinfo {pages} {061103} (\bibinfo {year} {2013})}\BibitemShut {NoStop}%
\bibitem [{\citenamefont {Bouzid}\ \emph {et~al.}(2015)\citenamefont {Bouzid}, \citenamefont {Massobrio}, \citenamefont {Boero},\ and\ \citenamefont {Ori}}]{Bouzid2015}%
  \BibitemOpen
  \bibfield  {author} {\bibinfo {author} {\bibfnamefont {A.}~\bibnamefont {Bouzid}}, \bibinfo {author} {\bibfnamefont {C.}~\bibnamefont {Massobrio}}, \bibinfo {author} {\bibfnamefont {M.}~\bibnamefont {Boero}},\ and\ \bibinfo {author} {\bibfnamefont {G.}~\bibnamefont {Ori}},\ }\href {https://doi.org/10.1103/PhysRevB.92.134208} {\bibfield  {journal} {\bibinfo  {journal} {Physical Review B}\ }\textbf {\bibinfo {volume} {92}},\ \bibinfo {pages} {134208} (\bibinfo {year} {2015})}\BibitemShut {NoStop}%
\bibitem [{\citenamefont {Schumacher}\ \emph {et~al.}(2016)\citenamefont {Schumacher}, \citenamefont {Weber}, \citenamefont {J{\'o}v{\'a}ri}, \citenamefont {Tsuchiya}, \citenamefont {Youngs}, \citenamefont {Kaban},\ and\ \citenamefont {Mazzarello}}]{Schumacher2016}%
  \BibitemOpen
  \bibfield  {author} {\bibinfo {author} {\bibfnamefont {M.}~\bibnamefont {Schumacher}}, \bibinfo {author} {\bibfnamefont {H.~H.}\ \bibnamefont {Weber}}, \bibinfo {author} {\bibfnamefont {P.}~\bibnamefont {J{\'o}v{\'a}ri}}, \bibinfo {author} {\bibfnamefont {Y.}~\bibnamefont {Tsuchiya}}, \bibinfo {author} {\bibfnamefont {T.~G.~A.}\ \bibnamefont {Youngs}}, \bibinfo {author} {\bibfnamefont {I.}~\bibnamefont {Kaban}},\ and\ \bibinfo {author} {\bibfnamefont {R.}~\bibnamefont {Mazzarello}},\ }\href {https://doi.org/10.1038/srep27434} {\bibfield  {journal} {\bibinfo  {journal} {Scientific Reports}\ }\textbf {\bibinfo {volume} {6}},\ \bibinfo {pages} {27434} (\bibinfo {year} {2016})}\BibitemShut {NoStop}%
\bibitem [{\citenamefont {Privitera}\ \emph {et~al.}(2016)\citenamefont {Privitera}, \citenamefont {Mio}, \citenamefont {Smecca}, \citenamefont {Alberti}, \citenamefont {Zhang}, \citenamefont {Mazzarello}, \citenamefont {Benke}, \citenamefont {Persch}, \citenamefont {La~Via},\ and\ \citenamefont {Rimini}}]{Privitera2016}%
  \BibitemOpen
  \bibfield  {author} {\bibinfo {author} {\bibfnamefont {S.~M.~S.}\ \bibnamefont {Privitera}}, \bibinfo {author} {\bibfnamefont {A.~M.}\ \bibnamefont {Mio}}, \bibinfo {author} {\bibfnamefont {E.}~\bibnamefont {Smecca}}, \bibinfo {author} {\bibfnamefont {A.}~\bibnamefont {Alberti}}, \bibinfo {author} {\bibfnamefont {W.}~\bibnamefont {Zhang}}, \bibinfo {author} {\bibfnamefont {R.}~\bibnamefont {Mazzarello}}, \bibinfo {author} {\bibfnamefont {J.}~\bibnamefont {Benke}}, \bibinfo {author} {\bibfnamefont {C.}~\bibnamefont {Persch}}, \bibinfo {author} {\bibfnamefont {F.}~\bibnamefont {La~Via}},\ and\ \bibinfo {author} {\bibfnamefont {E.}~\bibnamefont {Rimini}},\ }\href {https://doi.org/10.1103/PhysRevB.94.094103} {\bibfield  {journal} {\bibinfo  {journal} {Phys. Rev. B}\ }\textbf {\bibinfo {volume} {94}},\ \bibinfo {pages} {094103} (\bibinfo {year} {2016})}\BibitemShut {NoStop}%
\bibitem [{\citenamefont {Li}\ \emph {et~al.}(2018)\citenamefont {Li}, \citenamefont {Chen}, \citenamefont {Wang},\ and\ \citenamefont {Sun}}]{https://doi.org/10.1002/adfm.201803380}%
  \BibitemOpen
  \bibfield  {author} {\bibinfo {author} {\bibfnamefont {X.~B.}\ \bibnamefont {Li}}, \bibinfo {author} {\bibfnamefont {N.-K.}\ \bibnamefont {Chen}}, \bibinfo {author} {\bibfnamefont {X.-P.}\ \bibnamefont {Wang}},\ and\ \bibinfo {author} {\bibfnamefont {H.-B.}\ \bibnamefont {Sun}},\ }\href {https://doi.org/10.1002/adfm.201803380} {\bibfield  {journal} {\bibinfo  {journal} {Advanced Functional Materials}\ }\textbf {\bibinfo {volume} {28}},\ \bibinfo {pages} {1803380} (\bibinfo {year} {2018})}\BibitemShut {NoStop}%
\bibitem [{\citenamefont {Mukhopadhyay}\ \emph {et~al.}(2016)\citenamefont {Mukhopadhyay}, \citenamefont {Lindsay},\ and\ \citenamefont {Singh}}]{Mukhopadhyay2016}%
  \BibitemOpen
  \bibfield  {author} {\bibinfo {author} {\bibfnamefont {S.}~\bibnamefont {Mukhopadhyay}}, \bibinfo {author} {\bibfnamefont {L.}~\bibnamefont {Lindsay}},\ and\ \bibinfo {author} {\bibfnamefont {D.~J.}\ \bibnamefont {Singh}},\ }\href {https://doi.org/10.1038/srep37076} {\bibfield  {journal} {\bibinfo  {journal} {Scientific Reports}\ }\textbf {\bibinfo {volume} {6}},\ \bibinfo {pages} {37076} (\bibinfo {year} {2016})}\BibitemShut {NoStop}%
\bibitem [{\citenamefont {Sklénard}\ \emph {et~al.}(2021)\citenamefont {Sklénard}, \citenamefont {Triozon}, \citenamefont {Sabbione}, \citenamefont {Nistor}, \citenamefont {Frei}, \citenamefont {Navarro},\ and\ \citenamefont {Li}}]{sklenard_electronic_2021}%
  \BibitemOpen
  \bibfield  {author} {\bibinfo {author} {\bibfnamefont {B.}~\bibnamefont {Sklénard}}, \bibinfo {author} {\bibfnamefont {F.}~\bibnamefont {Triozon}}, \bibinfo {author} {\bibfnamefont {C.}~\bibnamefont {Sabbione}}, \bibinfo {author} {\bibfnamefont {L.}~\bibnamefont {Nistor}}, \bibinfo {author} {\bibfnamefont {M.}~\bibnamefont {Frei}}, \bibinfo {author} {\bibfnamefont {G.}~\bibnamefont {Navarro}},\ and\ \bibinfo {author} {\bibfnamefont {J.}~\bibnamefont {Li}},\ }\href {https://doi.org/10.1063/5.0073469} {\bibfield  {journal} {\bibinfo  {journal} {Applied Physics Letters}\ }\textbf {\bibinfo {volume} {119}},\ \bibinfo {pages} {201911} (\bibinfo {year} {2021})}\BibitemShut {NoStop}%
\bibitem [{\citenamefont {Yin}\ and\ \citenamefont {Chen}(2018)}]{yin_microstructure_2018}%
  \BibitemOpen
  \bibfield  {author} {\bibinfo {author} {\bibfnamefont {Q.}~\bibnamefont {Yin}}\ and\ \bibinfo {author} {\bibfnamefont {L.}~\bibnamefont {Chen}},\ }\href {https://doi.org/10.1007/s10854-018-9746-0} {\bibfield  {journal} {\bibinfo  {journal} {J Mater Sci: Mater Electron}\ }\textbf {\bibinfo {volume} {29}},\ \bibinfo {pages} {16523} (\bibinfo {year} {2018})}\BibitemShut {NoStop}%
\bibitem [{\citenamefont {Andrikopoulos}\ \emph {et~al.}(2007)\citenamefont {Andrikopoulos}, \citenamefont {Yannopoulos}, \citenamefont {Kolobov}, \citenamefont {Fons},\ and\ \citenamefont {Tominaga}}]{andrikopoulos_raman_2007}%
  \BibitemOpen
  \bibfield  {author} {\bibinfo {author} {\bibfnamefont {K.}~\bibnamefont {Andrikopoulos}}, \bibinfo {author} {\bibfnamefont {S.}~\bibnamefont {Yannopoulos}}, \bibinfo {author} {\bibfnamefont {A.}~\bibnamefont {Kolobov}}, \bibinfo {author} {\bibfnamefont {P.}~\bibnamefont {Fons}},\ and\ \bibinfo {author} {\bibfnamefont {J.}~\bibnamefont {Tominaga}},\ }\href {https://doi.org/10.1016/j.jpcs.2007.02.027} {\bibfield  {journal} {\bibinfo  {journal} {Journal of Physics and Chemistry of Solids}\ }\textbf {\bibinfo {volume} {68}},\ \bibinfo {pages} {1074} (\bibinfo {year} {2007})}\BibitemShut {NoStop}%
\bibitem [{\citenamefont {Wei}\ \emph {et~al.}(2017)\citenamefont {Wei}, \citenamefont {Wei}, \citenamefont {Zhang}, \citenamefont {Zhao},\ and\ \citenamefont {Zhang}}]{wei_grayscale_2017}%
  \BibitemOpen
  \bibfield  {author} {\bibinfo {author} {\bibfnamefont {T.}~\bibnamefont {Wei}}, \bibinfo {author} {\bibfnamefont {J.}~\bibnamefont {Wei}}, \bibinfo {author} {\bibfnamefont {K.}~\bibnamefont {Zhang}}, \bibinfo {author} {\bibfnamefont {H.}~\bibnamefont {Zhao}},\ and\ \bibinfo {author} {\bibfnamefont {L.}~\bibnamefont {Zhang}},\ }\href {https://doi.org/10.1038/srep42712} {\bibfield  {journal} {\bibinfo  {journal} {Sci Rep}\ }\textbf {\bibinfo {volume} {7}},\ \bibinfo {pages} {42712} (\bibinfo {year} {2017})}\BibitemShut {NoStop}%
\bibitem [{\citenamefont {Vinod}\ \emph {et~al.}(2015)\citenamefont {Vinod}, \citenamefont {Ramesh},\ and\ \citenamefont {Sangunni}}]{vinod_structural_2015}%
  \BibitemOpen
  \bibfield  {author} {\bibinfo {author} {\bibfnamefont {E.~M.}\ \bibnamefont {Vinod}}, \bibinfo {author} {\bibfnamefont {K.}~\bibnamefont {Ramesh}},\ and\ \bibinfo {author} {\bibfnamefont {K.~S.}\ \bibnamefont {Sangunni}},\ }\href {https://doi.org/10.1038/srep08050} {\bibfield  {journal} {\bibinfo  {journal} {Sci Rep}\ }\textbf {\bibinfo {volume} {5}},\ \bibinfo {pages} {8050} (\bibinfo {year} {2015})}\BibitemShut {NoStop}%
\bibitem [{\citenamefont {Kozyukhin}\ \emph {et~al.}(2013)\citenamefont {Kozyukhin}, \citenamefont {Veres}, \citenamefont {Nguyen}, \citenamefont {Ingram},\ and\ \citenamefont {Kudoyarova}}]{kozyukhin_structural_2013}%
  \BibitemOpen
  \bibfield  {author} {\bibinfo {author} {\bibfnamefont {S.}~\bibnamefont {Kozyukhin}}, \bibinfo {author} {\bibfnamefont {M.}~\bibnamefont {Veres}}, \bibinfo {author} {\bibfnamefont {H.}~\bibnamefont {Nguyen}}, \bibinfo {author} {\bibfnamefont {A.}~\bibnamefont {Ingram}},\ and\ \bibinfo {author} {\bibfnamefont {V.}~\bibnamefont {Kudoyarova}},\ }\href {https://doi.org/10.1016/j.phpro.2013.04.011} {\bibfield  {journal} {\bibinfo  {journal} {Physics Procedia}\ }\textbf {\bibinfo {volume} {44}},\ \bibinfo {pages} {82} (\bibinfo {year} {2013})}\BibitemShut {NoStop}%
\bibitem [{\citenamefont {Bala}\ and\ \citenamefont {Thakur}(2022)}]{bala_ga_2022}%
  \BibitemOpen
  \bibfield  {author} {\bibinfo {author} {\bibfnamefont {N.}~\bibnamefont {Bala}}\ and\ \bibinfo {author} {\bibfnamefont {A.}~\bibnamefont {Thakur}},\ }\href {https://doi.org/10.1007/s10854-022-08365-9} {\bibfield  {journal} {\bibinfo  {journal} {J Mater Sci: Mater Electron}\ }\textbf {\bibinfo {volume} {33}},\ \bibinfo {pages} {14419} (\bibinfo {year} {2022})}\BibitemShut {NoStop}%
\bibitem [{\citenamefont {Xu}\ \emph {et~al.}(2018)\citenamefont {Xu}, \citenamefont {Chen}, \citenamefont {Wang}, \citenamefont {Wu}, \citenamefont {Chong},\ and\ \citenamefont {Ye}}]{xu_optical_2018}%
  \BibitemOpen
  \bibfield  {author} {\bibinfo {author} {\bibfnamefont {Z.}~\bibnamefont {Xu}}, \bibinfo {author} {\bibfnamefont {C.}~\bibnamefont {Chen}}, \bibinfo {author} {\bibfnamefont {Z.}~\bibnamefont {Wang}}, \bibinfo {author} {\bibfnamefont {K.}~\bibnamefont {Wu}}, \bibinfo {author} {\bibfnamefont {H.}~\bibnamefont {Chong}},\ and\ \bibinfo {author} {\bibfnamefont {H.}~\bibnamefont {Ye}},\ }\href {https://doi.org/10.1039/C8RA01382A} {\bibfield  {journal} {\bibinfo  {journal} {{RSC} Adv.}\ }\textbf {\bibinfo {volume} {8}},\ \bibinfo {pages} {21040} (\bibinfo {year} {2018})}\BibitemShut {NoStop}%
\bibitem [{\citenamefont {Talochkin}\ \emph {et~al.}(2021)\citenamefont {Talochkin}, \citenamefont {Kokh},\ and\ \citenamefont {Tereshchenko}}]{talochkin_optical_2021}%
  \BibitemOpen
  \bibfield  {author} {\bibinfo {author} {\bibfnamefont {A.~B.}\ \bibnamefont {Talochkin}}, \bibinfo {author} {\bibfnamefont {K.~A.}\ \bibnamefont {Kokh}},\ and\ \bibinfo {author} {\bibfnamefont {O.~E.}\ \bibnamefont {Tereshchenko}},\ }\href {https://doi.org/10.1134/S002136402110012X} {\bibfield  {journal} {\bibinfo  {journal} {Jetp Lett.}\ }\textbf {\bibinfo {volume} {113}},\ \bibinfo {pages} {651} (\bibinfo {year} {2021})}\BibitemShut {NoStop}%
\bibitem [{\citenamefont {Mio}\ \emph {et~al.}(2017)\citenamefont {Mio}, \citenamefont {Privitera}, \citenamefont {Bragaglia}, \citenamefont {Arciprete}, \citenamefont {Cecchi}, \citenamefont {Litrico}, \citenamefont {Persch}, \citenamefont {Calarco},\ and\ \citenamefont {Rimini}}]{mio_role_2017}%
  \BibitemOpen
  \bibfield  {author} {\bibinfo {author} {\bibfnamefont {A.~M.}\ \bibnamefont {Mio}}, \bibinfo {author} {\bibfnamefont {S.~M.~S.}\ \bibnamefont {Privitera}}, \bibinfo {author} {\bibfnamefont {V.}~\bibnamefont {Bragaglia}}, \bibinfo {author} {\bibfnamefont {F.}~\bibnamefont {Arciprete}}, \bibinfo {author} {\bibfnamefont {S.}~\bibnamefont {Cecchi}}, \bibinfo {author} {\bibfnamefont {G.}~\bibnamefont {Litrico}}, \bibinfo {author} {\bibfnamefont {C.}~\bibnamefont {Persch}}, \bibinfo {author} {\bibfnamefont {R.}~\bibnamefont {Calarco}},\ and\ \bibinfo {author} {\bibfnamefont {E.}~\bibnamefont {Rimini}},\ }\href {https://doi.org/10.1038/s41598-017-02710-3} {\bibfield  {journal} {\bibinfo  {journal} {Sci Rep}\ }\textbf {\bibinfo {volume} {7}},\ \bibinfo {pages} {2616} (\bibinfo {year} {2017})}\BibitemShut {NoStop}%
\bibitem [{\citenamefont {Abou El~Kheir}\ \emph {et~al.}(2024)\citenamefont {Abou El~Kheir}, \citenamefont {Bonati}, \citenamefont {Parrinello},\ and\ \citenamefont {Bernasconi}}]{abou_el_kheir_unraveling_2024}%
  \BibitemOpen
  \bibfield  {author} {\bibinfo {author} {\bibfnamefont {O.}~\bibnamefont {Abou El~Kheir}}, \bibinfo {author} {\bibfnamefont {L.}~\bibnamefont {Bonati}}, \bibinfo {author} {\bibfnamefont {M.}~\bibnamefont {Parrinello}},\ and\ \bibinfo {author} {\bibfnamefont {M.}~\bibnamefont {Bernasconi}},\ }\href {https://doi.org/10.1038/s41524-024-01217-6} {\bibfield  {journal} {\bibinfo  {journal} {npj Comput Mater}\ }\textbf {\bibinfo {volume} {10}},\ \bibinfo {pages} {33} (\bibinfo {year} {2024})}\BibitemShut {NoStop}%
\bibitem [{\citenamefont {Sosso}\ \emph {et~al.}(2009)\citenamefont {Sosso}, \citenamefont {Caravati}, \citenamefont {Gatti}, \citenamefont {Assoni},\ and\ \citenamefont {Bernasconi}}]{sosso_vibrational_2009}%
  \BibitemOpen
  \bibfield  {author} {\bibinfo {author} {\bibfnamefont {G.~C.}\ \bibnamefont {Sosso}}, \bibinfo {author} {\bibfnamefont {S.}~\bibnamefont {Caravati}}, \bibinfo {author} {\bibfnamefont {C.}~\bibnamefont {Gatti}}, \bibinfo {author} {\bibfnamefont {S.}~\bibnamefont {Assoni}},\ and\ \bibinfo {author} {\bibfnamefont {M.}~\bibnamefont {Bernasconi}},\ }\href {https://doi.org/10.1088/0953-8984/21/24/245401} {\bibfield  {journal} {\bibinfo  {journal} {Journal of Physics: Condensed Matter}\ }\textbf {\bibinfo {volume} {21}},\ \bibinfo {pages} {245401} (\bibinfo {year} {2009})}\BibitemShut {NoStop}%
\end{thebibliography}%

\end{document}